\documentclass[11pt]{article}
\usepackage{times}
\usepackage{graphicx}
\usepackage{natbib}

\begin{document}

\title{Cosmological N-Body Simulations}

\author{J.S.Bagla and T.Padmanabhan \\
Inter-University Centre for Astronomy and Astrophysics, \\
Post Bag 4, Ganeshkhind, Pune 411 007, INDIA}

\date{ }

\maketitle

\begin{abstract}
In this review we discuss Cosmological N-Body codes with a special
emphasis on Particle Mesh codes.  We present the mathematical model
for each component of N-Body codes.  We compare alternative methods
for computing each quantity by calculating errors for each of the
components.  We suggest an optimum set of components that can be
combined reduce overall errors in N-Body codes.
\end{abstract}

\section{Introduction}

Observations suggest that the present universe is populated by very
large structures like galaxies, clusters of galaxies etc.  Current
models for formation of these structures are based on the assumption
that gravitational amplification of small perturbations leads to the
formation of large scale structures.  In absence of analytical
methods for computing quantities of interest, numerical simulations
are the only tool available for study of clustering in the non-linear
regime.  The last two decades have seen a rapid development of
techniques and computing power for cosmological simulations and the results
of these simulations have provided valuable insight into the study of
structure formation.  In this paper we will describe some aspects of
the cosmological N-body simulation, based on the code developed and
used by the authors.  We will stress the key physical and numerical
ideas and detail some useful tests of the N-Body code.  After an
initial introduction to cosmological N-Body codes the discussion of
N-Body codes will focus on Particle Mesh [PM] codes.  [For a
comprehensive discussion of numerical simulations using particles, see
Hockney and Eastwood (1988).] 

An N-body code consists of two basic modules: one part computes the
force field for a given configuration of particles and the other moves
the particles in this force field. These two are called at each step
to ensure that the force field and the particle trajectories evolve in
a self consistent manner.  Apart from these we also need to setup the
initial conditions and write the output.  Thus the basic plan of an
N-Body code follows the flow chart shown in figure~1.

Structure of these modules depends on the specific application at
hand.  Here we outline some features that are particular to
cosmological N-Body codes.  To make quantitative statements about the
required parameters of an N-Body simulation we shall assume that the
particles populate a region of volume $V$.  We shall also assume that
we are using $N_p$ particles to describe the density field.  The
following physical requirements have to be taken into account while
writing cosmological N-Body codes.

\begin{figure}
\begin{center}
\includegraphics[width=4 truein]{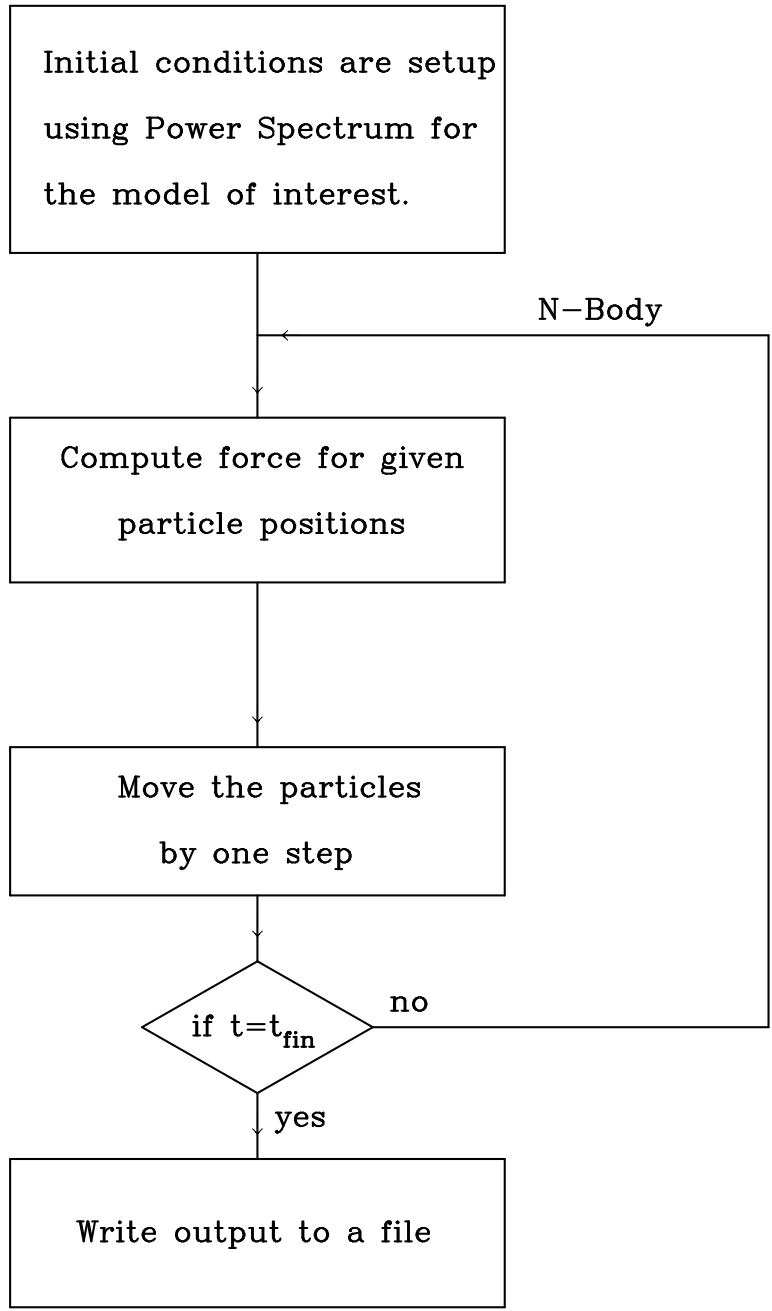}
\end{center}
\caption{Flow chart for an N-Body Code.}
\end{figure}

\begin{itemize}
\item  The universe is filled with matter over scales that are much
larger than the scales of interest in an N-Body simulation.  Density
averaged over larger and larger scales in the universe approaches a
constant value.  Thus the simulation volume $V$ can not be assumed to
exist in isolation and we must fill the region outside this volume by
some method.  Periodic boundary conditions are the only viable
solution to this problem.  In absence of periodic boundary conditions
most of the matter in the simulation volume has a tendency to collapse 
towards the centre of the box, giving
rise to a spurious, large rate of growth for perturbations.  [A compromise
solution, called quasiperiodic boundary conditions, has been tried to
solve this problem for tree codes.  In this scheme periodic boundary
conditions are used to restructure the simulation volume to bring a
particle at the centre of the box while computing force at its
position.  This is done for each particle so that there is no strong
tendency to collapse towards the centre of the box. (\cite{qperi})]
The most natural geometry for a volume with periodic boundary
conditions is a cube.

\item  We would like the evolution of perturbations to be independent
of the specific boundary conditions.  Thus the simulation volume should
not be dominated by any one object.  [If the simulation volume is
dominated by one object then the tidal force due to its own periodic
copies will influence its later evolution.]

\item  We require the average density in the box to equal the average
density of the universe.  Thus perturbations averaged at the scale of
the box must be ignorable at all times for the model of interest, i.e.,
$\sigma(R=V^{1/3}) \ll 1$.  [Here $\sigma(R)$ is the rms dispersion in the 
mass averaged at scale $R$.] 
For example, in case of the standard CDM model this would require the
box to be at least $100 h^{-1}$Mpc [at $z=0$] in extent.

\item  We would like to probe scales that are sufficiently non-linear
in order to justify the use of N-Body simulations.  If we wish to
compare the results with the real universe then we should be able to
study the formation of structures with a large range in scales.  In
other words, the mass of individual particles in an ideal simulation
must be less than the mass of the smallest structure of interest.
Therefore the smallest number of particles required to 
cover the relevant range of scales to study galaxy clustering, say, is 
\begin{equation}
N \geq \frac{M(100 h^{-1}{\rm Mpc})}{M_{galaxy}}  \simeq  \frac{5
\times 10^{18}M_\odot}{10^{11} M_\odot} \simeq 5 \times 10^7
\end{equation}
We need a very large number of particles to represent the density field
over this range of scales.  Therefore most numerical techniques used in
cosmological N-Body codes are oriented towards reducing the number of
operations and stored quantities per particle. 

\item We approximate collection of a very large number of particles in
the universe by one particle in an N-Body simulation.  Therefore the
particles in an N-Body simulation must interact in a purely collisionless
manner.
\end{itemize}
The periodic boundary conditions and a very large number of particles are
the key considerations in developing the detailed algorithm for an N-Body
code.  Of the two main components in an N-Body code, integration of the
equation of motion is a process of order $O(N_p)$.  The calculation of force,
if done by summing the force over all pairs, is a process of order
$O(N_p^2)$.  Therefore the calculation of force is likely to be more time
consuming than evolving the trajectories in N-Body codes with a large
number of particles.  It can be shown that, for particle numbers greater  
than a few hundred, direct computation of force takes an excessively long
time even on the fastest computers.  [The very high speed special purpose
computers are an exception to this. (\cite{grape})]  Three schemes have
been evolved to address this problem and replace direct summation over
pairs by methods that are less time consuming.  These are

\begin{itemize}
\item {\sl Particle Mesh} (PM) : Poisson equation is solved in the Fourier
domain and the potential/force is computed on a fixed grid.  The force is then
interpolated to particle positions for moving the particles.  Density
field, the source for gravitational potential, is also computed on the same
mesh/grid from particle positions by using a suitable interpolating
function.  The ``smoothing'' of particles limits the resolution of such
simulations but ensures collisionless evolution.  [See Bouchet, Adam and
Pellat (1984) and Bouchet and Kandrup, (1985) for an excellent discussion
of PM codes.]

\item {\sl Particle-Particle Particle Mesh} (P$^3$M) : This scheme
(Efstathiou et al, 1985) improves the resolution of PM method by adding a
correction to the mesh force for pairs of particles with separation of the
order of, and smaller than, the grid length.  The number of operations
required for this correction is proportional to $N_p n(R)$ where
$n(R)$ is the average number of neighbouring particles within a
distance $R$.  This can be written as $N_p \bar{n} ( 1 + \bar\xi(R))$,
where $\bar n$ is the average number density and $\bar\xi$ is the
averaged correlation function.  It is obvious that such corrections
can become time consuming in highly clustered situations [when
$\bar\xi(R) \gg 1$].

\item {\sl Tree} : In this scheme, information about the density field
is set up in form of a hierarchical tree.  Each level of the tree specifies
position of the centre of mass and the total mass for a region in the
simulation volume.  Force from particles in distant regions in the
simulation box is approximated by the force from the centre of mass for
particles in that region.  This leads to a reduction in the number of
operations required for calculating force.  [Barnes and Hut (1986);
Bouchet and Hernquist (1988)]  To estimate the number of
operations required for setting up the tree structure and evaluate the
force, let us assume that the simulation volume is a cube and is
subdivided into eight equal parts at each level.  This subdivision of
cells is continued till we have at most one particle per cell at the
smallest level.  Each particle is parsed at most once at each level,
therefore the upper bound on the total number of operations is
proportional to $N_p 
\ln(L_{box}/l_{min})$ where $l_{min}$ is the smallest inter-particle
separation for the given distribution of particles.  We have,
\begin{equation}
l_{min}\simeq n_{max}^{-1/3} = \delta_{max}^{-1/3} {\bar n}^{-1/3} =
N_p^{-1/3} L_{box} \delta_{max}^{-1/3} 
\end{equation}
where $\bar n$ is the average number density, $\delta_{max}$ is the
maximum density contrast and $n_{max}$ is the highest number density
in the given distribution of particles.  This implies that the upper
bound on number of operations is proportional to $O(N_p\ln{N_p
\delta_{max}})$.  Incorporating periodic boundary conditions is a very
nontrivial problem in the context of tree codes. [See Hernquist,
Bouchet and Suto (1991) for a discussion of periodic boundary
conditions for tree codes.] 
\end{itemize}

Amongst these methods the PM scheme has two advantages over the other
methods.  Firstly it is the only one of the three methods outlined above
that ensures a collisionless evolution.  [See Melott et al. (1997);
Suisalu and Saar (1996).]  Secondly it has the simplest 
algorithm and it is also the fastest of the three methods discussed above.  In
the remaining discussion we shall focus only on PM codes.  However, apart
from computation of force, all other components are the same in all these
codes [with some minor variations] and most of the conclusions about
relative merits of the available alternatives are applicable for all three
types of codes.

\section{Moving the Particles}

In this section we will first present the system of equations that
describe the evolution of trajectories of particles.  This will be
followed by a description of the methods used for numerical
integration of equation of motion.

\subsection{Equation of Motion}

It can be shown that the evolution of perturbations in a non-relativistic
medium in an expanding background can be studied in the Newtonian
limit at scales that are much smaller than the Hubble radius $d_H = c
H_0^{-1}$.  The equations for a set of particles interacting only
through the gravitational force can be written as
\begin{eqnarray}
& & {\bf \ddot x} + 2 \frac{\dot a}{a} {\bf \dot x} 
= - \frac{1}{a^2}\nabla_x\varphi 
 \nonumber \\
& & {} \nabla_x^2\varphi  
  = 4 \pi G a^2 \bar\varrho \delta
  = \frac{3}{2} H_0^2 \Omega_0 \frac{\delta}{a} .
 \label{dyn_st} 
\end{eqnarray}
Here the last equality follows from the Friedmann equations that
describe evolution of the scale factor for the universe.  The
variables used here are : comoving co-ordinates ${\bf x}$, time $t$,
scale factor $a$, gravitational potential due to perturbations
$\varphi$, density $\varrho$ and density contrast $\delta$.
Cosmological parameters used in this equation are : the Hubble's
constant $H_0$ and the density parameter contributed by
non-relativistic matter $\Omega_0$.

N-Body simulations integrate this equation numerically for each particle.
Numerical integration can be simplified by modifying the equation of motion
so that the velocity dependent term is removed from the equation of
motion. This can be achieved by using a different time parameter
$\theta$ (\cite{negtime}).  This is defined as
\begin{equation}
d\theta =  H_0 \frac{dt}{a^2} 
\end{equation}
In terms of which we have
\begin{eqnarray}
& & {} \frac{d^2 {\bf x}}{d \theta^2} = - \frac{3}{2} \Omega_0 a^2
\nabla\chi \nonumber \\ 
& & {} \nabla^2\chi = \frac{\delta}{a} \label{thetatime}
\end{eqnarray}
In this form, all the cosmological factors can be clubbed in the source term
in the equation of motion, making numerical implementation much simpler.

\subsection{Numerical Integration}

In any N-Body code, a great deal of computer time is devoted to integration
of the equation of motion.  The basic idea of time integration is simple :
the equation of motion expresses the second derivative of position in terms
of position, velocity and time.  Time integration translates this into the
subsequent changes in position and velocity.  In the following discussion
we will develop the idea of numerical integration of trajectories.

Writing the position and velocity at time $t+\epsilon$ in a Taylor series
about the position and velocity at time $t$, we have 
\begin{eqnarray}
& & x_i(t+\epsilon) = x_i(t) + \epsilon v_i(t) + \frac{1}{2}\epsilon^2
a_i(t)
 + \ldots \nonumber \\
& & v_i(t+\epsilon) = v_i(t) + \epsilon a_i(t) +
\frac{1}{2}\epsilon^2 j_i(t)  + \ldots 
\end{eqnarray}
Here $x$ is the position, $v$ is the velocity, $a$ is the acceleration
and $j$ is the rate of change of acceleration, and the subscript is used
to identify the particle.  We can use the equation of
motion to express acceleration and the higher derivatives of position in
terms of positions and velocities.  Therefore, in principle, we can find
the new position and velocity with arbitrary accuracy.  However, such a
scheme is not very practical from the computational point of view as it
requires an infinite series to be summed.  A more practical method consists
of truncating the series at some point and using sufficiently small time
steps to ensure the required level of accuracy. To illustrate this
point, let us consider terms up to first order in the above series, we get 
\begin{eqnarray}
& & {v_i}(t+\epsilon)={v_i}(t) + \epsilon a_i(t) + O(\epsilon^2)
\nonumber \\
& & {x_i}(t+\epsilon) = x_i(t) + \epsilon v_i(t) + O(\epsilon^2)  
\label{euler}
\end{eqnarray}
This is the Euler's method of solving differential equations and the error
in the solution is of order $\epsilon^2$.  Thus choosing a smaller
time step leads to smaller error in each step.  However, the number of
steps required to integrate over a given interval in time is inversely
proportional to the time step so that the overall error changes only linearly
with step size.

Let us consider terms up to $\epsilon^2$ in order to device a more accurate
integrator.
\begin{eqnarray}
& & x_i(t+\epsilon) = x_i(t) + \epsilon v_i(t) + \frac{1}{2}\epsilon^2
a_i(t)  + O(\epsilon^3) \nonumber \\
& & v_i(t+\epsilon) = v_i(t) + \epsilon a_i(t)+ \frac{1}{2}\epsilon^2 j_i(t)
+ O(\epsilon^3)
\end{eqnarray}
This equation contains the rate of change of acceleration and therefore is
not very useful in this form.  To simplify these equations without loosing
accuracy, let us specify $j$ as 
\begin{equation}
j(t) = \frac{a(t+\epsilon) - a(t)}{\epsilon} + O(\epsilon)
\end{equation}
This reduces the above equations to the following form
\begin{eqnarray}
& & x_i(t+\epsilon) = x_i(t) + \epsilon v_i(t)+ \frac{1}{2}\epsilon^2 a_i(t)
+ O(\epsilon^3) \nonumber \\
& & v_i(t+\epsilon) = v_i(t) +  \frac{\epsilon}{2}(a_i(t) + a_i(t+\epsilon))
 + O(\epsilon^3) \label{leapfrog}
\end{eqnarray}
This method can be used for velocity independent forces and is
identical to the Leap-Frog method.  The standard Leap-Frog method 
involves updating  the velocity and position at an offset of half a
step.  To see the equivalence of the above equations and the Leap-Frog
method let us consider the expressions for velocity at $t+\epsilon/2$
and $t-\epsilon/2$.  
\begin{eqnarray}
& & v_i(t+\epsilon/2) = v_i(t) + \frac{\epsilon}{2} a_i(t)+
\frac{1}{8}\epsilon^2 j_i(t) + O(\epsilon^3) \nonumber \\
& & v_i(t-\epsilon/2) = v_i(t) - \frac{\epsilon}{2} a_i(t)+
\frac{1}{8}\epsilon^2 j_i(t) + O(\epsilon^3)
\end{eqnarray}
These two equations can be combined to give
\begin{equation}
v_i(t+\epsilon/2) = v_i(t-\epsilon/2) + \epsilon a_i(t) +
O(\epsilon^3)
\end{equation}
We can also use the expression for velocity at $(t+\epsilon)$ to write
\begin{equation}
x_i(t+\epsilon) = x_i(t) + \epsilon v_i(t+\epsilon/2) + O(\epsilon^3)
\end{equation}
These two equations can now be combined to give the Leap-Frog method.
\begin{eqnarray}
& & x_i(t+\epsilon) = x_i(t) + \epsilon v_i(t+\epsilon/2) + 
O(\epsilon^3) \nonumber \\ 
& & v_i(t+3\epsilon/2) = v_i(t+\epsilon/2) + \epsilon
a_i(t) + O(\epsilon^3) \label{stleapfrog} 
\end{eqnarray}
This is called the Leap-Frog method as it updates velocities halfway
between the step that is used to update the position.  For velocity
dependent forces these methods have to be modified into a
predictor-corrector form.

\begin{table}
\begin{center}
\begin{tabular}{||l|l||} 
\hline
$\epsilon$ \ \ \ \ \ \ \ \ \  & $(\Delta
x)_{rms}$  \\ 
\hline
0.1 & 5.4 $\times$ $10^{-4}$ \\ 
0.05 &  7.3 $\times$ $10^{-5}$ \\ 
0.02 &  3.5 $\times$ $10^{-5}$ \\ 
0.01 &  1.2 $\times$ $10^{-5}$  \\
0.005 & 5.3 $\times$ $10^{-6}$  \\ 
\hline
\end{tabular}
\end{center}
\caption{This table lists the values of $\epsilon$ used in forward and
backward integration of trajectories of a set of particles in an
external potential and the corresponding error in recovering the
initial conditions.  Here we have used one grid length as the unit of
length.} 
\end{table}

It is possible to improve the accuracy by using integrators that retain
more terms in the Taylor series expansion.  However most of these schemes
require many evaluations of force for each step making them
computationally uneconomical. 

\subsection{Choosing the Time Step}

The time step for integrating the equation of motion can be chosen in many
ways.  In all of these methods one defines a suitable criterion [like energy
conservation] and then obtains the largest time step that will satisfy the
requirement. Following is a list of possible criteria : 
\begin{itemize}
\item Monitoring validity of Irvine-Layzer equation.  This is done by
integrating the Irvine-Layzer equation [Irvine (1961); Layzer (1963);
Dmitriev and Zel'dovich (1964)]
\begin{equation}
T + U + \int \frac{da}{a} \left( 2T + U \right) = C
\end{equation}
and monitoring the constant $C$.  Variation in $C$ with respect to total
energy $T+U$ can be used as an indicator of non conservation. [This is
normally done with different forms of this equation so that only the
kinetic or the potential energy term contributes to the
integral. (\cite{p3m})]  In schemes where the force is computed at mesh
points and then interpolated to particle positions the force at the
particle position, ${\bf f}({\bf x}) \neq -\nabla_{\bf x} \varphi$,
i.e., the force does not equal the gradient of potential at a point. 
This leads to an apparent non conservation of energy.  If
we include a correction term in the energy equation to account for this
fact (\cite{p3m}) then it can be shown that it is possible to improve
conservation of energy.  This correction involves a direct sum over all
pairs of particles and is therefore an almost impossible task to implement
for a simulation with a large number of particles.  Thus this is not a
practical method of deciding the time step.

\item  Convergence of final positions and velocities.  If we evolve the
trajectories of a set of particles with different time steps, then we
expect the final positions and velocities to converge towards the correct
value as we reduce the time step.  This can be used to decide the time
step.

\item  Reproducibility of initial conditions.  This is a more stringent
version of the test outlined above.  This method ensures that
the solution we get is correct and errors are not building up
systematically.  If we run the particles forward and then back again
we should in principle 
get back the initial positions.  Although we ignore the decaying mode while
integrating trajectories forward in time, it tends to dominate if we evolve
the system in the opposite direction.  We can overcome this difficulty by
not evolving the potential during this test.
\end{itemize}

We find the last test to be the most stringent one and we use it to fix the
time step for an arbitrary potential by the following construct.
Consider an arbitrary potential $\psi$ which has the maximum value of
the magnitude of its gradient as $g_{max} = |\nabla\psi|_{max}$.  We
can write the equation of motion eqn.(\ref{thetatime}) for a small
interval $\Delta\theta$ as 
\begin{equation}
\frac{\epsilon}{(\Delta\theta)^2} = \frac{24}{\theta^4} g_{max}
\end{equation}
Here we have used $\epsilon$ to represent the largest displacement in
time $\Delta\theta$.  We have specialised to $\Omega_0 = 1$ in writing
the above equation.  In which case, $\theta=-2/\sqrt{a}$. For a given
value of $\epsilon$ we can fix the time step as
\begin{equation}
\Delta\theta = \theta^2 \sqrt{\frac{\epsilon}{24 g_{max}}}
\end{equation}
We can use the test mentioned above to fix the value of $\epsilon$.
Table~1 lists a few values of $\epsilon$ and $(\Delta x)_{rms}$ for
a CDM simulation [boxsize=$64 h^{-1}$Mpc] with $64^3$ particles in a
$64^3$ box.  The system was evolved from $a=0.02$ to $0.25$ and then
back to $a=0.02$.  We use $\epsilon=0.005$ for our simulations as we
find the error in this case to be acceptable. 

\section{Calculating the Force}

An N-Body code solves the equation of motion and the Poisson equation
in a self consistent manner.  Therefore, moving the particles and
computing the force for a given distribution of particles are the two
most important components of an N-Body code.  In this section we will
discuss computation of force at mesh points for a given distribution
of particles.  For simplicity we shall assume that all the particles have
the same mass.  Generalisation to particles with different masses is not
difficult but can strain the resources like RAM.  [Setting up initial
conditions for simulations that have particles with different masses
is much simpler than setting up initial conditions for simulations
that use particles with equal mass. See \S{4} for details.] 

In particle mesh codes the force at the particle position can be
computed from the potential defined at mesh points in two ways.  These
are :
\begin{itemize}
\item {\sl Energy Conserving Schemes} : The force is computed at
particle positions by using the gradient of the interpolating function.
\begin{equation}
{\bf f}({\bf x}) = -\nabla_{\bf x}\left( \sum\limits_{{\bf
x}_i} W({\bf x},{\bf x}_i) \psi({\bf x}_i)\right)
\end{equation}
The sum is over all the mesh points.  An additional requirement is
that the same 
interpolating function $W$ should be used to compute the density field on
the mesh from the distribution of particles.  These are called energy
conserving schemes as the force and gravitational potential are
related to each other as ${\bf f}({\bf x}) = -\nabla_{\bf x} \psi({\bf
x})$.

\item {\sl Momentum Conserving Schemes} : The force is computed on the
mesh and then interpolated to the particle position.
\begin{equation}
{\bf f}({\bf x}) = -\sum\limits_{{\bf x}_i} W({\bf x},{\bf x}_i)
\nabla_{{\bf x}_i}\psi({\bf x}_i)
\end{equation}
Momentum conserving schemes also require the same 
interpolating function to be used for computing density field on
the mesh.  It can be shown that in these schemes the force due to the
$i$th particle 
on the $j$th particle is same as the force due to $j$th particle on
the $i$th particle.  [${\bf f}_{ij} = - {\bf f}_{ji}$.]  This clearly
is a necessary condition for momentum conservation.
\end{itemize}
Use of mesh leads to anisotropy in force at scales comparable to the
mesh size.  A simple method for limiting these anisotropies is to
choose a configuration that leads to a softer force at the mesh
scale.  Collisionless evolution also requires the force to drop rapidly
below the average inter-particle separation (\cite{plane}).  We choose
to work with the momentum conserving schemes as these lead to a softer
force at small scales in comparison with the energy conserving schemes.

The algorithm for computing force at particle positions involves the
following major steps.
\begin{itemize}
\item  {\sl Computing the Density Field} : Mass of particles is
assigned to mesh points by using an interpolating function.  Density
contrast is calculated from this ``mass field'' by using the definition
\begin{equation}
\delta = \frac{M}{\bar M} - 1 \label{delta}
\end{equation}
Here $\bar M$ is the average mass per mesh point. Then fast Fourier
transform technique is used to compute the Fourier components of
density contrast.

\item  {\sl Solving the Poisson Equation} : The Poisson equation is solved
in Fourier space.  The FFT can be used to compute the gravitational
potential on the mesh.

\item  {\sl Computing the Force} :  The gravitational force equals the
negative gradient of the gravitational potential, therefore the next
step in algorithm is computing the gradient of potential.  This can be
done either in Fourier space or in real space by using some scheme for
numerical differentiation.  Force at the particle positions is computed by
interpolation from the mesh points.
\end{itemize}
In the following subsections we will discuss each of these steps in
greater detail.

\subsection{Computing the Density Contrast}

Smoothing [interpolating] functions are used to assign mass to the
lattice points for a given configuration of particles.  This is done
by computing the following sum for every mesh point.
\begin{equation}
M({\bf x}_i) = \sum\limits_{\hbox{all particles}} M_{particle} W({\bf
x},{\bf x}_i) 
\end{equation}
Here ${\bf x}_i$ are the co-ordinates of mesh points, ${\bf x}$ are the
coordinates of particles and $W$ is the interpolating function.  As we
have assigned same mass to all the particles we can scale the mass at
mesh cites with the particle mass.

Density contrast is computed from the mass defined at mesh points by
using eqn.(\ref{delta}).  FFT is then used to compute its Fourier
transform.

The smoothing function $W$ should satisfy certain algebraic,
computational and physical requirements.
\begin{itemize}
\item The sum of weights given to lattice points should equal unity.
The interpolating function should be continuous and preferably
differentiable as well. 

\item The smoothing function should be peaked at small separations and
should decay rapidly at large separations.  The smoothing function
should have the smallest possible base, i.e., the number of lattice
separations over which it is non-zero should be small.  The number of
arithmetic operations required for interpolation is proportional to
the cube of the number of lattice cells over which the interpolating
function is non-zero.

\item The interpolating function should be isotropic.
\end{itemize}
It is clear that we can not satisfy all these criteria fully and at
the same time and therefore we have to consider a compromise solution.
The first simplification that is made is to break up the three
dimensional interpolating function into a product of three one
dimensional interpolating functions.  This reduces the
complexity of the problem but also implies that the interpolating
function cannot be manifestly isotropic.

These considerations, and the above mentioned simplifying assumption
then leads to the following set of  interpolating functions.
\begin{itemize}
\item  {\sl Nearest Grid Point} [NGP] : This is the simplest
interpolating function.  In this assignment scheme all the mass 
is assigned to the nearest grid point. 
\begin{equation}
W(x,x_i)=\left\{ \hbox{ } \begin{array}{ll}
1 \hbox{ \ \ } & \hbox{(for $|x-x_i| \leq L/2$)} \\
0  \hbox{ \ \ } & \hbox{(for $|x-x_i| > L/2$)} \\
\end{array} \right.
\end{equation}
where $L$ is the grid spacing. This clearly satisfies the algebraic
condition that the sum of weights should equal unity.  However, this
function is neither continuous nor differentiable. The leading term in
the error in force of a point mass computed using this mass assignment
scheme has a dipole like behaviour.

\begin{figure}
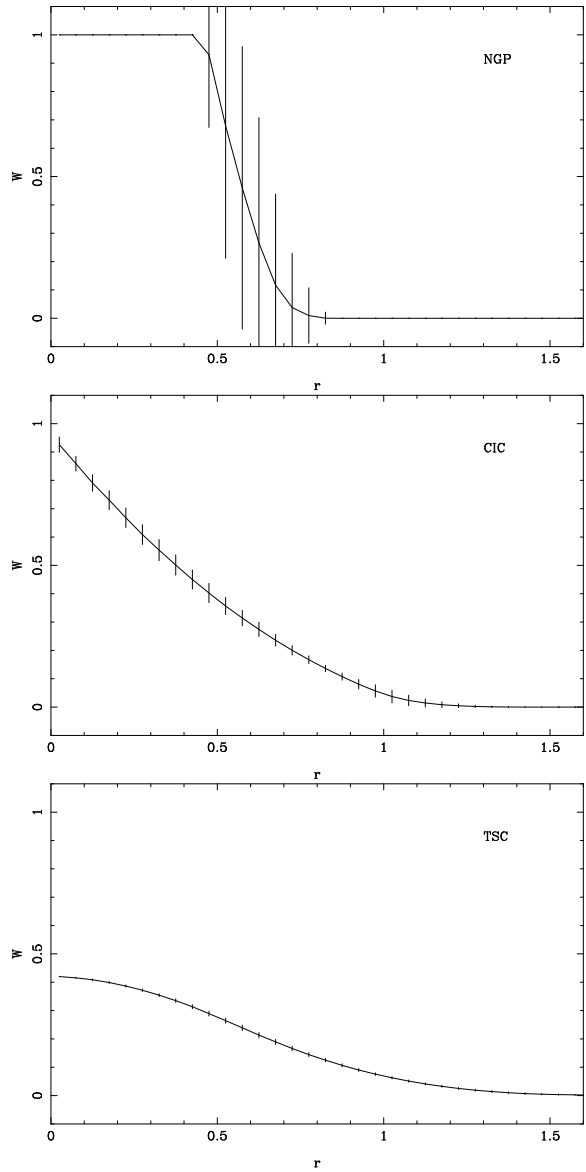

\begin{center}
\includegraphics[width=3truein]{disp_ngp.ps}
\includegraphics[width=3truein]{disp_cic.ps}
\includegraphics[width=3truein]{disp_tsc.ps}
\end{center}
\caption{This figure shows the weight assigned to a point that is at
a distance $r$ from the mesh point of interest.  The curve in each of
the panels shows the average weight where the average is computed over
all directions [$\theta,\phi$],  keeping $r$ fixed.  The vertical
bars show the {\it rms} dispersion about this average to depict the
level of anisotropies in these functions.}
\end{figure}

\item  {\sl Cloud In Cell} [CIC] : This the most commonly used interpolating
function.  Linear weights are used and mass is assigned
to the two nearest grid points.  The function is continuous but not
differentiable.
\begin{equation}
W(x,x_i)=\left\{ \hbox{ } \begin{array}{ll}
1 - |x-x_i|/L \hbox{ \ \ } & \hbox{ (for $|x-x_i| \leq L$)} \\
0 \hbox{ \ \ } & \hbox{ (for $|x-x_i| > L$)} \\
\end{array} \right.
\end{equation}

\item  {\sl Triangular Shaped Cloud} [TSC] : The third function we
wish to discuss is a quadratic spline that is continuous and also has
a continuous first derivative. In this case the mass is distributed
over three lattice points in each direction.
\begin{equation}
W(x,x_i)=\left\{ \hbox{ } \begin{array}{ll}
{3\over 4} - \left({\Delta x\over L}\right)^2 \hbox{ } & \hbox{(for
$\Delta x \leq L/2$)} \\
{9\over 8}\left( 1 - \left|{2\Delta x\over 3L}\right|\right)^2 \hbox{ } &
\hbox{(for $L/2 \leq \Delta x \leq 3L/2$)} \\
0 \hbox{ } & \hbox{(for $3L/2 \leq \Delta x$)} \\
\end{array} \right.
\end{equation}
Here $\Delta x=|x-x_i|$ is the magnitude of separation between the particle
and the mesh point.  This function satisfies the requirements from the
physical point of view but it is generally considered expensive from
the computational point of view.
\end{itemize}  
It is obvious that these smoothing functions become more and more
computationally intensive as we go for a larger base.  All of the
functions described above satisfy
the essential algebraic requirements, so that the final choice is
a compromise between the accuracy of the force field obtained using
these and the computational requirements.  However, we do not go beyond
TSC as the base becomes very large and the transverse error in force
becomes unacceptably large.

Figure 2 compares these smoothing functions in a realistic
situation.  We have plotted the average weight assigned to a point that is at
a distance $r$ from the mesh point of interest.  The curve in each of
the panels shows the average weight where the average is computed over
all directions [$\theta,\phi$] while keeping $r$ fixed.  The vertical
bars show the {\it rms} dispersion about this average to depict the
level of anisotropies in each of these functions.  It is clear that
TSC is the most isotropic of these functions and CIC is a
``reasonable'' compromise between isotropy and computational 
requirements.  We will use the TSC smoothing function in the N-Body
code for all applications.

\subsection{Solving the Poisson Equation}

The equation we wish to solve is
\begin{equation}
\nabla^2 \psi = \frac{\delta}{a}
\end{equation}
This equation is to be solved with periodic boundary conditions.  In
particle mesh codes the equation is solved in the Fourier space.  To
solve this equation with periodic boundary conditions we must ensure
that only the harmonics of the fundamental wave number $k_f = 2\pi/NL$
contribute to the function.  [$NL$ is the size of the simulation box.]
This type of sampling of $k$ space is
also required by the fast Fourier transforms.  Therefore as long as we
use FFT for computing Fourier transforms the correct boundary
conditions are automatically incorporated.

Two types of Green's functions are commonly used for solving the
Poisson equation in Fourier space.  These are

\begin{figure}
\begin{center}
\includegraphics[width=4.5 truein]{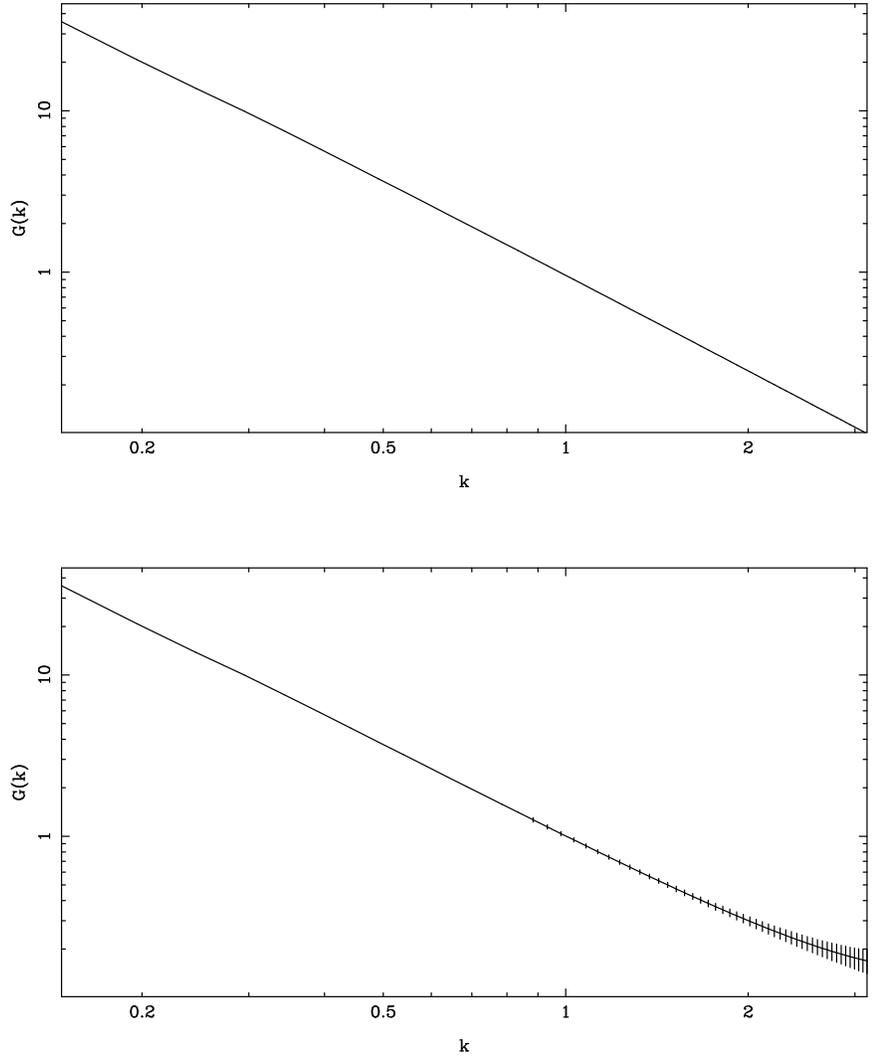}
\end{center}
\caption{This figure shows the two Green's functions described in the
text.  Upper panel shows the continuum Green's function and the lower
panel shows the Green's function derived from discrete representation
of the Laplacian.  Curves in each panel show the average of $G(k)$
over angular coordinates for any value of $k=|{\bf k}|$.  The vertical
bars show the {\it rms} dispersion about this average and are an
indicator of anisotropy.}
\end{figure}
\begin{itemize}
\item  {\sl Poor Man's Poisson Solver} : This method uses the
continuum Green's function.
\begin{eqnarray}
& & \psi_k = \frac{\delta_k}{a} G(k) \nonumber \\
& & G(k)=-\frac{1}{k^2} 
\end{eqnarray}
The main advantage of this is that it is the true
Green's function.  Also, it is isotropic by construction.

\item  {\sl Periodic Green's Function} : An alternative is to
substitute Fourier transform of a discrete realisation of the
Laplacian. A Green's function derived in this manner tends to increase
power at small scales and leads to larger anisotropies in the force
field.  A commonly used Green's function of this class is
\begin{equation}
G(k) = - \frac{L^2}{4 \left[\sin^2{k_xL/2} +\sin^2{k_yL/2} +
\sin^2{k_zL/2}\right]}
\end{equation}
Here $L$ is the grid spacing and equals unity in the units that we are
using here.  This Green's function is invariant
under the transformation $k_x \rightarrow  k_x + 2\pi$.  All FFT
routines use such a transformation to rearrange wave modes for ease of
manipulation [See \S{3}].  Therefore this Green's function is
easier to implement as compared to the continuum version as one does
not have to worry about the rearrangement of wave modes.  The main
disadvantage of this Green's function is that it is anisotropic at
small scales. 
\end{itemize}
\begin{figure}
\begin{center}
\includegraphics[width=5.5 truein]{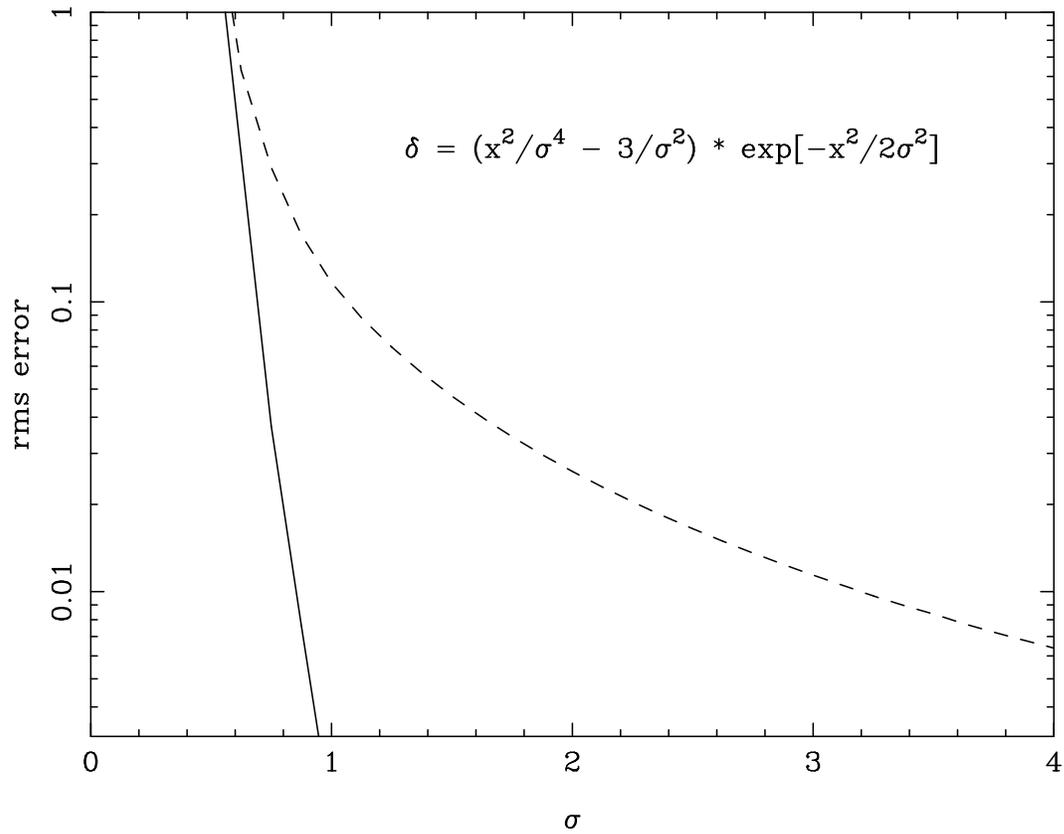}
\end{center}
\caption{This figure shows the {\it rms} error in solving Poisson
equation as a function of the parameter $\sigma$ defined in the text.
The dashed line shows the error for the periodic kernel whereas the
thick line shows the error for the poor man's Poisson solver. The
comparison was carried out on the grid so as to avoid mixing errors
in solving the equation with errors introduced by interpolation etc.}
\end{figure}

We have plotted the two Green's functions in figure~3.  Like
figure~2 the curve here shows the average over angular coordinates
for any value of $k=|{\bf k}|$.  The vertical bars show the {\it rms}
dispersion about this average and are an indicator of anisotropy. It is
clear from this figure that the periodic Green's function enhances
power at small scales, beyond the true value, and also introduces some
anisotropy in the potential. 

We compared these two Green's functions by solving the Poisson
equation for the density contrast
\begin{equation}
\delta = A \left( \frac{r^2}{\sigma^4} - \frac{3}{\sigma^2} \right)
\exp\left[ -\frac{r^2}{2\sigma^2}\right]
\end{equation}
The solution to the Poisson equation for this source can be obtained
analytically and the potential is
\begin{equation}
\psi = \frac{A}{a} \exp\left[ -\frac{r^2}{2\sigma^2}\right] + {\rm const.}
\end{equation}

We solved the Poisson equation numerically using the two Green's
functions defined above on a $64^3$ mesh.  We computed the {\it rms}
deviation from the true solution in each case for a large range of
values for $\sigma$.  The result of this analysis is presented in
figure~4 that shows the error as a function of $\sigma$.  This curve
clearly shows that the periodic kernel introduces more error
than the poor man's Poisson solver.

\subsection{Computing the Gradient of Gravitational Potential}

Numerical differentiation is used to compute the force [$-\nabla\psi$] on
the mesh.  The gravitational potential $\psi$ is also defined
on the mesh.  Interpolation is used to compute the force at the
particle position.

Here we will discuss three most commonly used methods for computing the
gradient of potential in N-Body codes.

\begin{figure}
\begin{center}
\includegraphics[width=5truein]{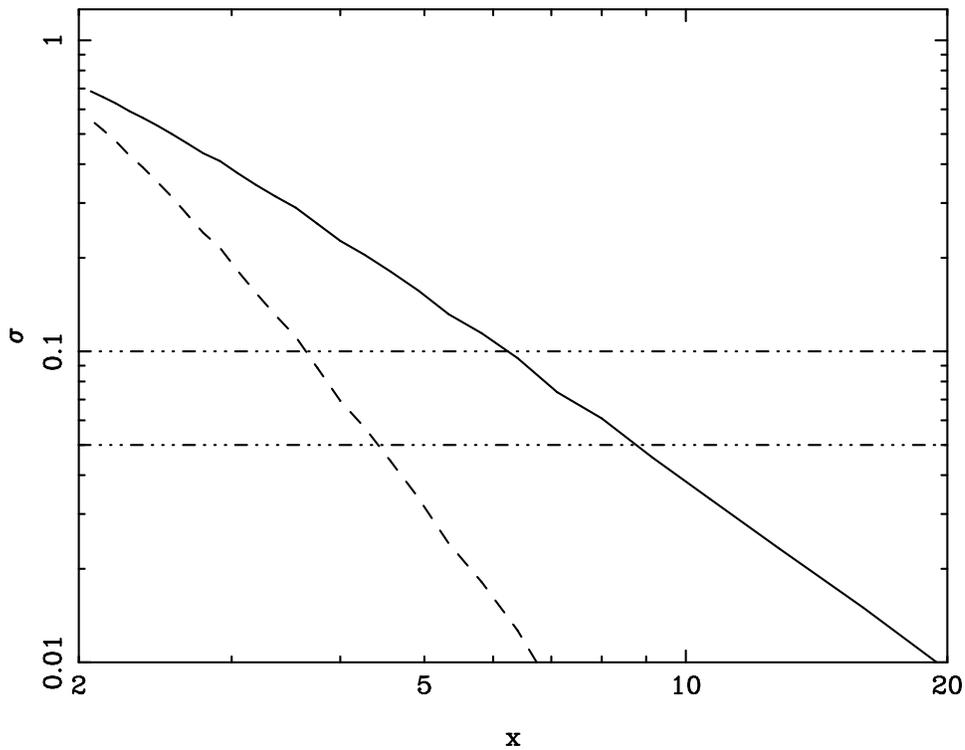}
\end{center}
\caption{This figure shows {\it rms} error in numerical differentiation.
We differentiated sine waves with frequencies $k_x$, $k_y$ and $k_z$ where
$k_i = 2\pi/x_i$.  We then compared the numerical result with its
analytical counterpart to obtain an estimate of error.  We have plotted
the error as a function of $x = 2\pi/(k_x^2 + k_y^2 + k_z^2)^{1/2}$. Dashed
line shows error for the five point formula and the thick line shows
error for the mid point rule.  Horizontal dot-dashed lines show the $10\%$ and
the $5\%$ error level.  This error analysis was carried out in a $64^3$
box.}
\end{figure}
\begin{itemize}
\item  The most accurate and fastest method for computing derivatives uses
fast Fourier transforms.  The Fourier components of force are computed
directly from the Fourier components of gravitational potential.  Only
possible source of errors here is the FFT routine itself.  The FFT routine
used here gives an error of about $10^{-4} \%$ for Fourier transforming a
sine wave in a $64^3$ box.  [This error was estimated by comparing
numerical values of Fourier components with their expected value.]

The expression for Fourier transform of a derivative is
\begin{equation}
{f'}_k = - i k f_k
\end{equation}

\item  The simplest method for computing the derivative in real space is
the mid point rule.
\begin{equation}
\frac{\partial f}{\partial x} = \frac{f(x+h) - f(x-h)}{2h} + O(h^2)
\end{equation}
The equivalent expression for computing the derivative on the mesh is
\begin{equation}
\frac{\partial f}{\partial n} = \frac{f(n+1) - f(n-1)}{2}
\end{equation}
For an individual sine wave, the fractional error varies as $k^2$ for
small $k$.  Here, as for every operation in cosmological simulations, 
periodic boundary conditions are used to compute the derivative near 
the edges.

\item  It is possible to construct a more accurate formula by using the
method of undetermined coefficients [See for example Antia (1991)].  A
five point formula 
for differentiation that gives fractional errors proportional to $k^4$ for
small $k$ is 
\begin{eqnarray}
\frac{\partial f}{\partial x} &=& \frac{f(x-2h) - f(x+2h) + 8(f(x+h) -
f(x-h))}{12h} \nonumber\\
& & \hbox{ \ \ \ \ \ }  + O(h^5)
\end{eqnarray}
This translates into the following expression for a uniformly spaced grid
:  
\begin{equation}
\frac{\partial f}{\partial n} = \frac{f(n-2) - f(n+2) + 8(f(n+1) -
f(n-1))}{12}
\end{equation}
\end{itemize}
To compare these methods we studied the variation of relative error for
sine waves as a function of frequency.  We computed the numerical
derivatives of a set of sine waves in a $64^3$ box and compared these with
the analytical results.  We have plotted the root mean square error as a
function of the inverse of wavenumber $x=2\pi/k$ in figure~5.  Dashed
line shows error for the five point formula and the thick line
corresponds to the mid point rule.  It is clear from these curves that the
five point formula is far more accurate than the mid point rule.

We have listed in table~2 the wavenumber for at which the average
error crosses some thresholds for these two formulae. 
\begin{table}
\begin{center}
\begin{tabular}{||l|l|l||}
\hline
Method \hbox{ \ \ \ \ \ \ } & $5\%$ error $(k,l)$ \hbox{ \ \ \ }  & $10\%$
error $(k,l)$ \hbox{ \ \ \ } \\
\hline
Mid-Point & $0.7$, $9.0$ & $1.0$, $6.3$ \\
\hline
Five Point & $1.4$, $4.5$ & $1.7$, $3.7$ \\
\hline
\end{tabular}
\end{center}
\caption{This table lists the scale at which error in differentiation
crosses the $5\%$ and $10\%$ level for the mid-point and the five
point methods.  A 
comparison of these values suggests that the five point method is
more robust and less error prone of the two.}
\end{table}

\subsection{Tests of Force Calculation}

In the preceding subsections we have studied the steps involved
in computing the force field for a given distribution of particles.
For each step we discussed alternate methods of computing the
required quantity and also discussed their relative merits.  We
shall now discuss the errors in computing the force field for one
particle.  We will use many combinations of individual components like
the interpolating function, Green's function and the differentiator and
compare the errors for each combination.

\begin{figure}
\begin{center}
\includegraphics[width=5truein]{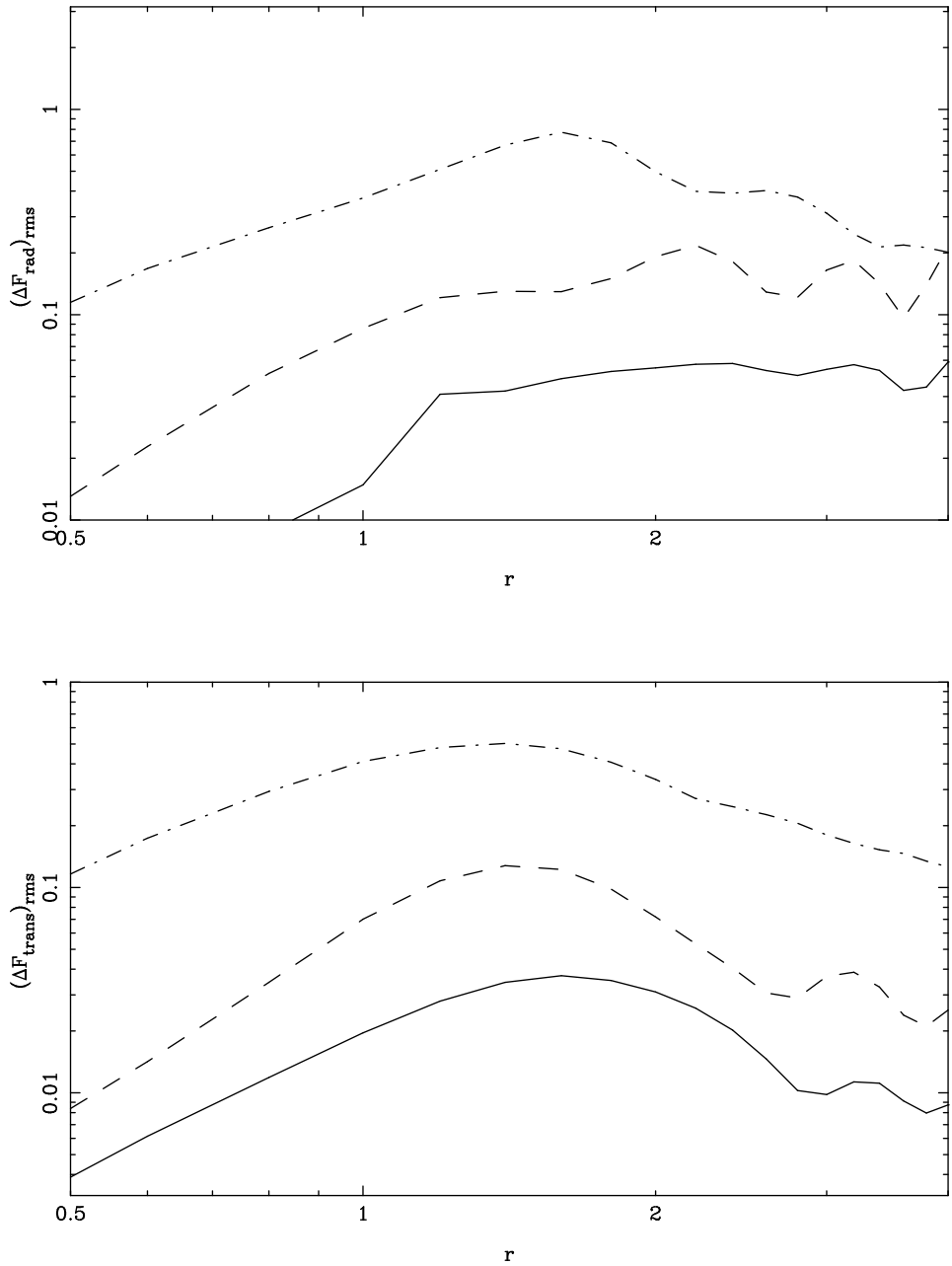}
\end{center}
\caption{This figure shows the error in force of one particle.  The
top panel shows the root mean square dispersion in the 
radial force.  The lower panel shows corresponding curves for the
transverse component of force.  The thick line shows the error for TSC
interpolating function, the dashed line for CIC and the dot-dashed
line corresponds to NGP.  In these panels the potential was
computed using the poor man's Poisson solver and the gradient of the
potential was computed in the Fourier space.}
\end{figure}
\setcounter{figure}{5}
\begin{figure}
\begin{center}
\includegraphics[width=5truein]{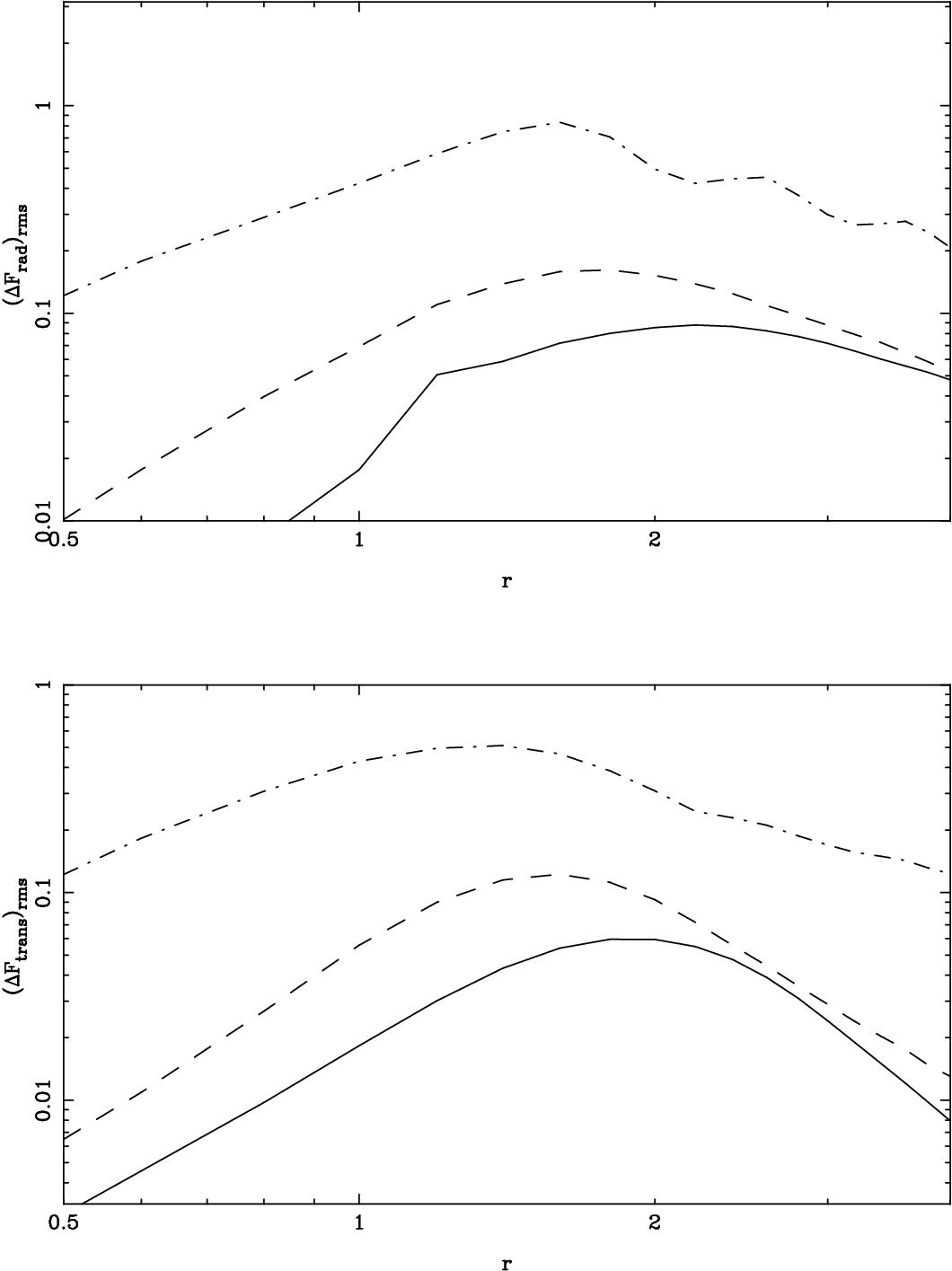}
\end{center}
\caption{Continued. In these panels the Poisson
equation was solved 
using the periodic Green's function and the mid point formula was used
for computing the force.  Notice that the CIC and TSC curves converge
in this case indicating that the error is dominated by some other
component.} 
\end{figure}

To compute the error we placed a particle at a random position in a
given lattice cell and computed force due to this at a fixed distance $r$ in
many directions to obtain the average force. We also computed the rms
dispersion about the average.  This process was repeated for one
hundred positions of the particle in the lattice cell.  Errors in this
force were computed by comparing with the true $1/r^2$ force expected
in such a situation.  The error was split into two parts : transverse
and radial.  The results of this study may be summarised as follows :
\begin{itemize}
\item  The average force deviates very little from the true force for
distances larger than one grid length [$r > L$] for almost all
combinations of components.  

\item  The root mean square dispersion about the average can be large for some
combinations of components.

\item  Smallest {\it rms} dispersion is obtained for the combination
of TSC, poor man's Poisson solver and numerical differentiation in
Fourier space.

\item  The combination of CIC, periodic kernel and mid point
differentiation also gives small errors.  Surprisingly the errors in
CIC and TSC [other components remaining same] {\it converge} at scales
larger than the grid size.  This clearly indicates that the error is
dominated by some other component, probably the mid-point
differentiation.  [Convergence disappears when five point formula or
differentiation in Fourier space is used.]
\end{itemize}
To elucidate some of the points made above we have plotted the radial
and transverse error for the two
combinations of components mentioned in the last two points.
Figure~6 shows that the error is minimum for TSC amongst the three
interpolating functions in all the cases.  Other points mentioned in
the above discussion are also brought out clearly in these panels.  We
will use the combination of TSC interpolating function and the poor man's
Poisson solver.  We will compute the gradient of potential in the Fourier
space.

\section{Fast Fourier Transforms}

In previous sections we discussed methods used for moving the
particles  and the steps involved in computing the force in N-Body
simulations.  We mentioned that fast Fourier transforms are 
used for computing the Fourier transforms and play a vital role in
reducing the time taken in computing the force.  In this section we
will start with the usual expression for the Fourier transform and
reduce it to the form in which it is processed by the FFT.  For
details of the FFT techniques we refer the reader to any standard book
on numerical analysis. (\cite{antia}; \cite{numrec})

For simplicity we will work in one dimension in this section.  However
we will write the $d$ dimensional generalisation of the main result.
Consider a function $f(x)$ that has the Fourier transform $g(k)$. Then
we may write 
\begin{equation}
f(x)=\int\limits_{-\infty}^{\infty}{\frac{dk}{2\pi} g(k) e^{-i kx}}
\end{equation}
It is not possible to use computers to integrate arbitrary functions
over an infinite range in finite time.  The integral is thus truncated
at some finite $k_{max}$.  This truncation is possible only for those
functions for which the integral over $|k|>k_{max}$ is sufficiently
small.  Thus the integral to be evaluated becomes
\begin{equation}
f(x)\simeq\int\limits_{-k_{max}}^{k_{max}}{\frac{dk}{2\pi} g(k) e^{-i
kx}}
\end{equation}
This integral is evaluated in form of a summation over
the integrand.
\begin{equation}
f(x)\simeq \sum\limits_{k=-k_{max}}^{k_{max}} g(k) e^{-ikx}
\frac{\Delta{k}}{2\pi} 
\end{equation}
where the function is sampled at intervals of $\Delta k$ in
this interval.  For a constant $\Delta k$, this
sampling will divide the range $-k_{max}$ to $k_{max}$ in $N$
intervals where $N=(2k_{max}/\Delta{k}) - 1$.  The fast Fourier transforms
require the number of intervals to equal an integer power of $2$,
which clearly is not possible for the above range if we are to sample
the $k=0$ mode as well.  It is possible to circumvent this problem by
making the range of integration asymmetric.  We add one extra interval
after $k_{max}$ and thus make the number of intervals equal to an even
number.  We can further choose $k_{max}$ and $\Delta k$ to ensure that
$N$ is an integer power of two.  Thus the sum may be written as
\begin{eqnarray}
f(x)&\simeq&{1\over NL}\sum_{-k_{max}}^{k_{max}+\Delta{k}}g(k)
e^{-ikx}\nonumber\\
&=&\frac{1}{NL}\sum_{m=-N/2+1}^{N/2} g(\Delta{k}\; m)
e^{-i\Delta{k}mx}
\end{eqnarray}
The boxsize $NL = 2\pi/\Delta k$ defines the scale over which the
function $f(x)$ is defined apart from its periodic copies.  Here
$N=(2k_{max}+\Delta k)/\Delta k = 2^j$ where $j$ is a positive
integer.  The function $f(x)$ is defined on a regular mesh with
$L$ as the spacing between adjacent mesh points.  This allows us to
rewrite the previous equation as
\begin{equation}
f(nL)={1\over{NL}}\sum_{m=-N/2+1}^{N/2} g(\Delta{k}\; m) e^{-i 2\pi
mn/N}
\end{equation}
To rewrite this in the form used by the FFT algorithm we split
this sum into two parts :
\begin{eqnarray}
f(nL)&=&\frac{1}{NL} \sum_{m=-N/2+1}^{-1} g(\Delta{k}\; m) e^{-i 2\pi
mn/N} \nonumber\\
& & \hbox{ } + \frac{1}{NL}\sum_{m=0}^{N/2} g(\Delta{k}\; m) e^{-i 2\pi
mn/N}
\end{eqnarray}
Changing the summation index in the first sum by $N$ [$m\rightarrow
m+N$] and dropping a $2\pi$ factor in the phase, we obtain
\begin{eqnarray}
f(nL)&=&\frac{1}{NL}\sum_{m=N/2+1}^{N-1} g(\Delta{k}(m-N)) e^{-i 2\pi
mn/N} \nonumber \\
& & \hbox{ } + \frac{1}{NL}\sum_{m=0}^{N/2} g(\Delta{k}\; m) e^{-i 2\pi mn/N}
\end{eqnarray}
These terms can be put together in the following form
\begin{equation}
f(nL)=\frac{1}{NL}\sum_{m=0}^{N-1} G(\Delta{k}\; m)
e^{-i 2\pi mn/N} \label{fft}
\end{equation}
where $G(\Delta{k}\; m)=g(\Delta{k}\; m)$ for $m\leq N/2$ and
$G(\Delta{k}\; m)= g(\Delta{k}(m-N))$ for $m > N/2$. From here it is
clear that the input array for FFT should have the negative wave
numbers {\it after} the positive wavenumbers.  [This can also be interpreted
as ``periodicity'' in $k$ space.]  The generalisation of the above
expression to $d$ dimensions is 
\begin{equation}
f({\bf n}L)=\frac{1}{(NL)^d}\sum_{{\bf m}={\bf 0}}^{\bf N-1}
G(\Delta{k}{\bf m}) e^{-i 2\pi {\bf m}.{\bf n}/N}
\end{equation} 
Here $G({\bf m})$ is the Fourier transform of the function $f({\bf
n}L)$ with every component $m_i > N/2$ replaced by $ m_i - N$.

If an FFT routine uses a different normalisation then the input array must
be rescaled to get the correct amplitude for the function $f({\bf
n}L)$. If the FFT routine uses
\begin{equation}
f({\bf n}L)=\frac{1}{l^d}\sum_{{\bf m}={\bf 0}}^{\bf N-1}
G(\Delta{k}{\bf m}) e^{-i 2\pi {\bf m}.{\bf n}/N} \label{scale}
\end{equation}
then we must scale the input function as $G^{FFT}=G*(l/{NL})^d$.  This
scaling is not important if we are computing both the forward and the
inverse transform as the overall normalisation must be same for all
FFT routines.  This scaling becomes relevant only in those cases where we are
transforming a given quantity only once, as in the generation of
initial potential.

The above equations establish the form in which an array must be
passed to the FFT routines in order to compute the Fourier transform.

\section{Setting up Initial Conditions}

\subsection{The Initial Density and Velocity Field}

N-Body simulations are generally started from fairly homogeneous
initial conditions, i.e. the density contrast is much smaller than
unity at all scales that can be studied using the simulation. In this
regime we can use linear theory to compute all quantities of
interest.

In linear theory, the evolution of density contrast can be described
as a combination of a growing and a decaying mode.  We can write the
solution for perturbations in an Einstein-deSitter Universe as
\begin{equation}
\delta(a)=c_1 a + c_2 a^{-3/2}
\end{equation}
We can evaluate $c_1$ and $c_2$ by using the initial conditions.
These conditions can be written in terms of the initial velocity field
and the initial potential.
\begin{eqnarray}
& & \delta_{in}= a_{in}\nabla^2\psi_{in} \nonumber \\
& & \left(\frac{\partial \delta}{\partial a}\right)_{in} = - \nabla
. {\bf u}_{in}
\end{eqnarray}
Using these we can express the linear solution in terms of the initial
potential and the initial velocity field as
\begin{equation}
\delta(a) = \left[ \frac{3}{5} \nabla^2 \psi_{in} - \frac{2}{5} {\bf
\nabla}.{\bf u}_{in}\right] a + \frac{2}{5} \left[ \nabla^2 \psi_{in}
+ {\bf \nabla}.{\bf u}_{in}\right] a^{-3/2}
\end{equation}
This equation suggests that for a system in growing mode, ${\bf
u}_{in} = -{\bf \nabla}\psi_{in}$. 
This equation gives us the density contrast at each point in terms of
the initial gravitational potential.  [We will discuss the problem of
generating the initial potential in the next subsection.]  Given an
initial potential the above equation suggests two schemes for
generating the initial density field.  These are :
\begin{itemize}
\item  The particle are distributed uniformly and their masses are
proportional to the local value of density.  We can either
start with zero velocities, in which case we have to increase the
amplitude of $\psi_{in}$ by a factor $5/3$ to account for the presence
of decaying mode.  Alternatively we can choose to put the system in
the growing mode and assign velocity ${\bf u}_{in} = -{\bf
\nabla}\psi_{in}$ to each particle.  One major drawback of this method
is that an extra array containing masses of particles needs to be
stored and this can be a problem for large simulations.

\item  Starting with a uniform distribution the particles are
displaced by a small amount, say much less than the 
inter-particle separation, using velocity ${\bf u}_{in} = -{\bf
\nabla}\psi_{in}$.  The resulting distribution of particles will
represent the required density field if the initial distribution did
not have any inhomogeneities.  We can also retain the initial velocity
field. 
\end{itemize}
Thus the main requirement from initial positions of particles before
we generate the required perturbation is that the distribution of
particles sample the potential ``uniformly.''  Any inhomogeneities
present in the initial distribution will combine with the density
perturbations that are generated by displacing particles and will
modify the initial power spectrum.

One obvious solution that ensures a uniform distribution is placing
particles on the grid.  This generates a distribution that is uniform
but not random.  An additional 
problem with this distribution is the extreme regularity which leads
to shot noise at Nyquist frequency for potentials with large coherence
length when the number of particles is smaller than the number of
cells in the mesh. [See figure~7]

Another ``obvious'' solution is to put particles at random inside the
simulation box.  This distribution is uniform but it has $\sqrt N$
fluctuations which result in spurious clustering.  The fluctuations in
number of particles in a cell of size $R$ decrease as $1/R^{3/2}$ with
scale.  [Comparing with power law spectra we note that this is an
$n=0$ spectrum.] 

Apart from these simple minded solutions there exist [at least] two
other distributions of particles which may be used for the initial 
positions. 

The {\it Glass} initial conditions are obtained by running a random
distribution of particles through N-Body with a repulsive force.  In
this case we start with small perturbations and the repulsive force
leads to a slow decrease in the amplitude of perturbations.  As the
perturbations get progressively smaller it worthwhile to estimate the
evolution of inhomogeneities using linear theory.  The equation for
evolution of perturbations in this case is same as the standard
equation for linear growth of perturbations (\cite{lssu}) 
but with a source term that has an opposite sign.  This leads to the
following oscillatory solution with a decaying amplitude.
\begin{equation}
\delta(a)= a^{-1/4} \left[ \alpha \cos\left({\sqrt{23} \over 4}
\ln{a}\right) + \beta \sin\left({\sqrt{23} \over 4} \ln{a}\right) \right]
\end{equation}
Where $\alpha$ and $\beta$ are constants to be fixed by initial
conditions.  The process of generating initial conditions can take a
long time for a large number of particles.  However this has to be
done only once and the data can be stored in a file for repeated use.

The other method is a variant of the grid and Poisson initial positions
mentioned above.  Here particles are placed in lattice cells but at a
random point within the cell.  This removes the regularity of grid
without sacrificing uniformity.  The number fluctuation falls faster
than $\sqrt{N}$ and have a smaller amplitude even at the smallest
scale.  The amplitude of fluctuations can be controlled by reducing the
amplitude of displacement about the mesh point.

After the initial density field has been generated the initial
velocities for N-Body can be set as ${\bf u}_{in} = -\nabla\psi_{in}$
if we want the system to be in the growing 
mode. However, generation of density field from the initial potential,
as outlined above involves many numerical operations that modify the
potential at scales comparable to the mesh size.  A better method, for
consistency of density, velocity and potential, is to recompute the
potential after generating the density field and set the velocities
with the recomputed potential (\cite{p3m}).  This is particularly
important in case of models with a lot of small scale power.  If we
use the initial potential for fixing velocities then velocities have a
stronger small scale component than would be expected from the density
field that has been generated, leading to inconsistency in the input
configuration for N-Body.  Therefore we use ${\bf u}_{in} = -{\bf
\nabla}\psi$, where $\nabla^2\psi = \delta(a) / a$ is the potential
generated from displaced distribution of particles.

\subsection{Initial Potential}

In this section we shall outline the method used to generate the
initial potential.  The same method can also be used to compute the
density field directly if the initial density field is to be generated
by placing particles with different masses on the mesh.

In most models of structure formation the initial density field is
assumed to be a Gaussian random field.  Linear evolution does not
modify the statistics of density fields except for evolving the
amplitude.  As the potential and density contrast are related through a
linear equation, it follows that the gravitational potential $\psi$ is
also a Gaussian random field at early epochs.  

A Gaussian random field is completely described in terms of its power
spectrum.  The Fourier components of a Gaussian random field [both the
real and the imaginary part] are random numbers with a normal
distribution  with variance proportional to the power spectrum of 
the random field.  The proportionality constant depends on the Fourier
transform convention.  To fix this constant we consider the probability
functional for a function $g$.  
\begin{equation}
{\rm P}[g] = B \exp\left[-\frac{1}{2} \int \frac{d^3{\bf k}}{(2\pi)^3}
\frac{\vert g(k)\vert^2}{P_g(k)}\right]
\end{equation}
Here $B$ is a normalisation constant.  In the Fourier transform
convention used in \S{3} this equation can be written as
\begin{equation}
{\rm P}[g] = B \exp\left[-{1\over 2} {1\over (NL)^3} \sum\limits_{all \,
{\bf k}} {\vert g(k)\vert^2 \over P_g(k)}\right]
\end{equation}
Thus the variance of the real and imaginary components for each
Fourier mode of a Gaussian random field should equal
\begin{equation}
\langle\vert g_{real}(k)\vert^2 \rangle  = \langle\vert
g_{imag}(k)\vert^2 \rangle  = {P_g(k) N^3 L^3 \over 2} 
\end{equation}
The factor of 2 takes into account the fact that both the real and the
imaginary components share the variance. Thus the function $g$ may be
written as
\begin{equation}
g(k)=(a_k + i b_k) \left[{P_g(k) N^3 L^3 \over 2}\right]^{1/2}
\end{equation}
where $a_k$ and $b_k$ are Gaussian random numbers with zero mean and
unit variance.

To generate the gravitational potential we substitute the
gravitational potential $\psi_k$ in place of $g(k)$ and the power
spectrum with the power spectrum for the gravitational potential.  The
power spectrum for the gravitational potential is $P_{\psi}(k,a) =
P_{\delta}(k,a)/(a^2k^4) = P_{\delta}^{lin}(a=1,k)/k^4$.  Here
$P_{\delta}^{lin}(a=1,k)$ is the linearly extrapolated power spectrum
of density fluctuations.   With this, we can write
\begin{equation}
\psi_k = (a_k + i b_k) \left[{P_{\delta}^{lin}(a=1,k) N^3 L^3 \over 2 k^4
}\right]^{1/2} .
\end{equation}
Here we have taken $a(t_0)=1$.  To specialise to the Fourier transform
convention used in the FFT routine we use, we use
$l=N^{1/2}$. [eqn.(\ref{scale})] This implies
\begin{equation}
{\psi_k}^{FFT} = { \psi_k \over N^{3/2} L^3} =  (a_k + i b_k)
\left[{P_{\delta}(k) \over 2 L^3 k^4 }\right]^{1/2}
\end{equation} 
Here $k=2\pi m/(NL)$.  If $P_\delta(k)$ is of the form $A k^n f(k)$
where $f$ is a dimensionless function [It may be the transfer
function or a cutoff imposed by hand.] and $A$ is the amplitude.  In
the units we are using here $L=1$, therefore 
\begin{equation}
{\psi_m}^{FFT} = (a_m + i b_m)\left[ {A\over 2} \left({ 2 \pi m
\over N}\right)^{n -4} f\left({ 2 \pi m \over N}\right) \right]^{1/2}
\end{equation}
where all lengths are written in units of $L$.

\subsection{Testing Initial Conditions}

In this section we shall test the initial conditions generated by the
method outlined above.  We will compare the power spectrum of density
perturbations generated by this method with the theoretical power
spectrum for a wide range of models.  This comparison was carried out
using simulations done with 
$128^3$ particles in a $128^3$ box.  To study the effect of a smaller
number of particles we also studied the power spectrum for a
simulation with $64^3$ particles in a $128^3$ box.  We use the
following models for this purpose :

\begin{figure}
\begin{center}
\includegraphics[width=5truein]{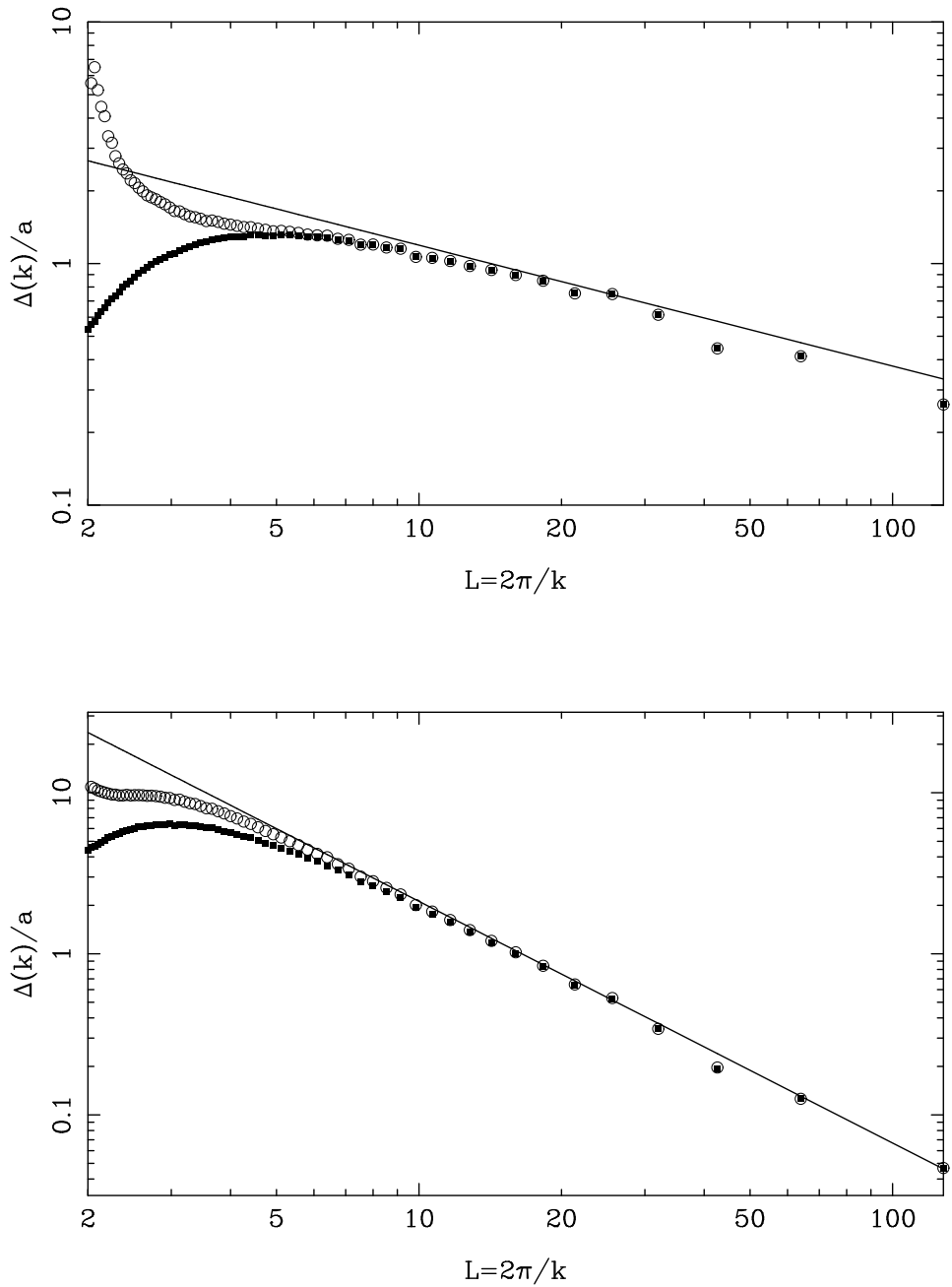}
\end{center}
\caption{This figure shows the theoretical power spectrum and the
induced power spectra, linearly extrapolated to $a=1$.  The
theoretical power spectrum is shown as a thick line and the power
spectrum generated by displacing the particles is shown as filled
squares [for simulations with $128^3$ particles] and as empty circles
[for simulations with $64^3$ particles].  The upper panel here
corresponds to the power law model with $n=-2$ and lower panel to the
model with $n=0$.}
\end{figure}
\setcounter{figure}{6}
\begin{figure}
\begin{center}
\includegraphics[width=5truein]{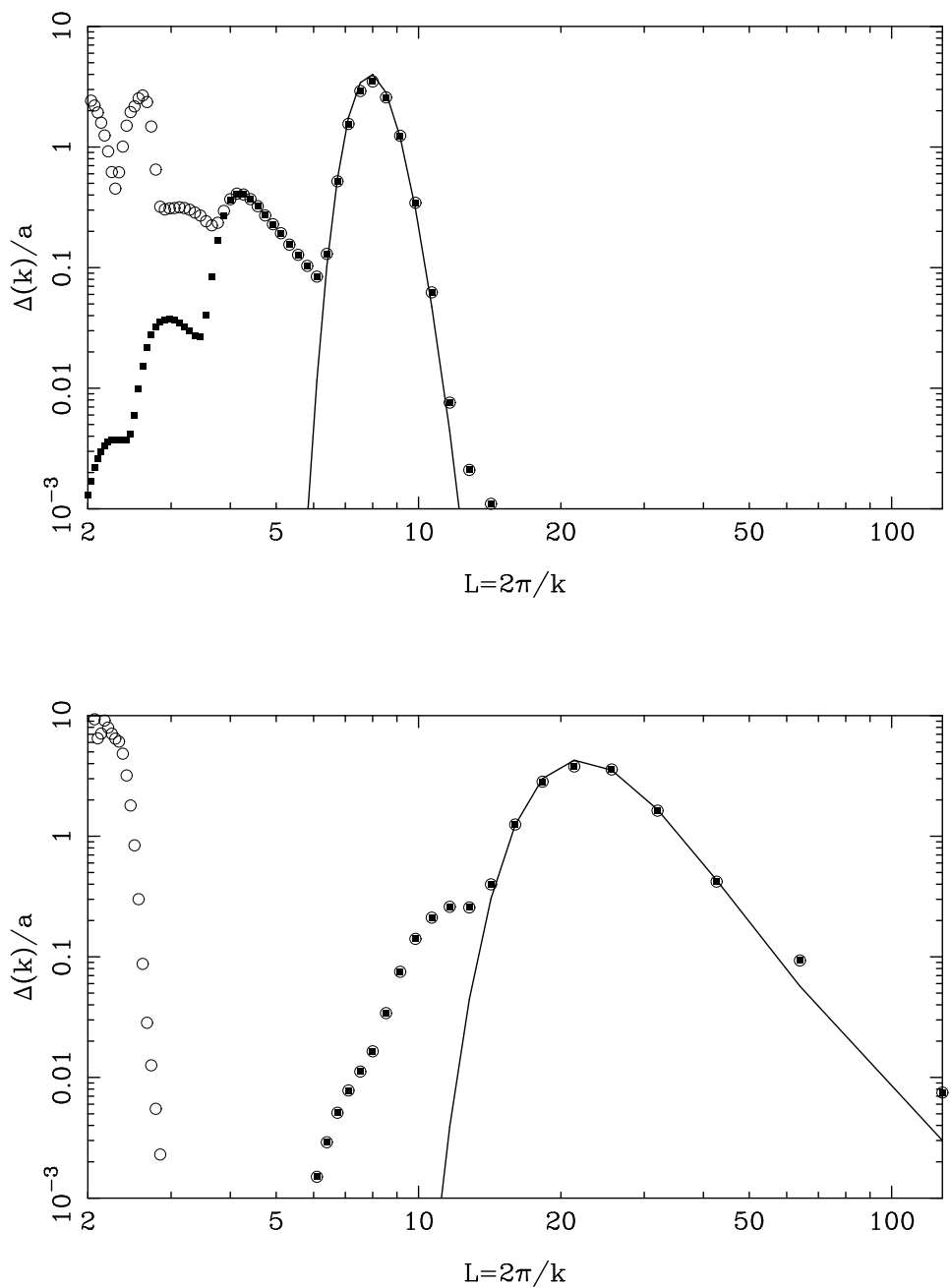}
\end{center}
\caption{Continued.  These panels show the initial power spectra for
models with Gaussian power spectrum.  The upper panel corresponds to a
Gaussian power spectrum with peak at $L_0=8L$ and the lower panel
corresponds to a spectrum with $L_0=24L$.  The secondary peaks at
small scales occur at the harmonics of the wavenumber corresponding to
the peak of the initial Gaussian.}
\end{figure}
\begin{itemize}
\item  Power law models with $n=-2$ and $n=0$.

\item  A Gaussian power spectrum with spread $\sigma_k = 2\pi/NL$
peaked at $k_0=2\pi/L_0$.  [$P(k) \propto
\exp\{-(k-k_0)^2/2\sigma_k^2\}$]  This model was studied for $L_0=8$ and $24$.
\end{itemize}
Models outlined in the first item have power at all scales whereas for
the Gaussian power spectrum the power is concentrated in a narrow
range of scales.  This tests the accuracy of initial condition
generation for the two extreme classes of models. 

We have plotted the theoretical power spectrum and the induced power
spectra, linearly extrapolated to $a=1$, in figure~7 for the models
listed above.  The theoretical power spectrum is shown as a thick line
and the power spectrum generated by displacing the particles is shown
as filled squares for simulations with $128^3$ particles and as
empty circles for simulations with $64^3$ particles.  

The simulated power spectrum agrees with the theoretical power
spectrum for scales larger than about $5L$ for power law models.  In
this range the power spectrum in simulations with different number of
particles also agrees quite well.  

In case of Gaussian power spectrum the simulated power spectra also
contain some power at harmonics of the main peak, i.e., at scales
$L_0/2$, $L_0/3$, etc.  The Gaussian at scale $L_0$ is reproduced
quite well.  Simulations with $64^3$ particles contain spurious noise
around the Nyquist scales.  Apart from this difference power spectra
in two cases agree quite well.

Simulations with $64^3$ particles generally have more noise at smaller
scales.  For hierarchical models this is essentially shot noise.
However for models with large scale coherence [as in case of Gaussian 
power spectrum with $L_0=24L$ or $n=-2$ power law model]  this can
lead to large spurious noise at the Nyquist scale.  This can be
understood in terms of the large coherence length in the displacement
field for such models, which leads to a sequence of alternately filled
and empty cells.  This contributes to the excess noise at the Nyquist
scale. 

The theoretical power spectrum is reproduced correctly in a
large range of scales for a wide variety of models.  Power spectrum at
scales smaller than about $5L$ does not match with the theoretical
power spectrum and hence results in this regime should not be used at
early times.  This constraint applies to all types of N-Body codes as
the same method is used to generate the initial conditions for all
types of N-Body codes.  This constraint does not apply at late times
as the power at small scales at late times is dictated by initial
power at larger scales. (\cite{crit_index})

\section{Tests of N-Body Code}

In this section we will test performance of the N-Body code as a
whole.  We will compare the evolution of gravitational clustering in
N-Body simulations with analytical results.  The number of analytical
results available in the non-linear regime is very limited, and hence
the number of tests is also very small.

As a test of evolution of positions and velocities we can compare the
motion for a one dimensional collapse before shell crossing with
Zel'dovich approximation (\cite{za}).  This test was suggested by 
Efstathiou et al.(1985) and they compared the positions and velocities
for such a collapse with the corresponding quantities in Zel'dovich
approximation.  They chose a sine wave along one of the coordinate
axes as the initial perturbation in potential.  Recently
(\cite{plane}) this test has been generalised so that the plane wave
is not along any coordinate axis.  In such a case the particles can
approach each other with a small nonzero impact parameter.  Monitoring
velocities along directions perpendicular to the perturbation provides a
good test of collisionless evolution.  Collisionless evolution of a
one dimensional perturbation can not induce velocities along
directions orthogonal to the perturbation.  The authors find that
Particle mesh codes are the only ones that pass this test.  P$^3$M and
tree codes can also provide collisionless evolution if the softening
length is of the same order as the inter particle separation.

We carried out this test with one dimensional perturbation along the
$x$ axis.  The results are listed in table~3.  Here we have tabulated
the root mean square error in displacement and velocity.  These are
defined here as
\begin{eqnarray}
& & \sigma_x = \left[\frac{\sum\limits_i \left(x_i -x_i^Z
\right)^2}{\sum\limits_i \left(x_i^Z -q_i \right)^2}\right]^{1/2} \nonumber \\
& & \sigma_v = \left[\frac{\sum\limits_i \left(v_i -v_i^Z
\right)^2}{\sum\limits_i \left(v_i^Z\right)^2}\right]^{1/2}
\end{eqnarray}
The table lists errors for three wavenumbers.  In each case we have
given the error at the epoch of first shell crossing in the system.  The
error increases with the wavenumber of the sine wave used for initial
potential as we sample a given wave with a smaller number of particles
and the discretisation effects become more important.

\begin{table}
\begin{center}
\begin{tabular}{||l|l|l||}
\hline
$k/k_f$ \hbox{ \ \ \ \ \ } & $\sigma_x$ \hbox{ \ \ \ \ \ } & $\sigma_v$ \\
\hline
1 & 0.021 & 0.034 \\
2 & 0.048 & 0.073 \\
4 & 0.104 & 0.138 \\
\hline
\end{tabular}
\end{center}
\caption{This table lists the errors in positions and velocities for
one dimensional collapse.  The errors are computed by comparing with
Zel'dovich approximation.  The errors have been listed for three
different initial perturbations, the first column contains the
wavenumber[in units of the fundamental wavenumber $k_f=2\pi/N$] of the
sine wave perturbation.  The errors were computed at the time of first
shell crossing in the system.  The N-Body simulation used for this was
carried out using $64^3$ particles.}
\end{table}

The above test checks the working of an N-Body code for a very special
case.  In a more general situation it is difficult to compare the
output of numerical simulation at the level of positions of particles
and only a statistical comparison is possible.  We will use the
averaged correlation function $\bar\xi$ and the scaled pair velocity
$h$ (\cite{lssu}) for testing the accuracy of the N-Body code.

\begin{figure}
\begin{center}
\includegraphics[width=5truein]{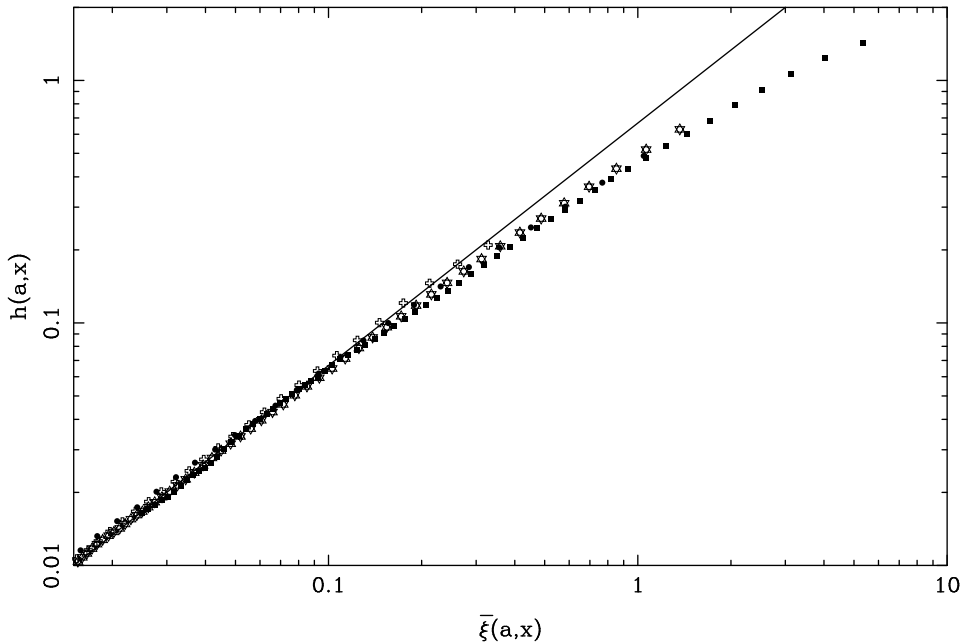}
\end{center}
\caption{This figure shows $h(a,x)$ as a function of $\bar\xi(a,x)$
for two power law models [$n=0$ and $-1$] and the CDM model.  The
thick line depicts the relation between these quantities in linear
theory and the results from simulations are shown as points.  All the
points in the range $\bar\xi \ll 1$ lie along the line with little
dispersion.}
\end{figure}

In linear regime the averaged correlation function $\bar\xi$ evolves
as $a^2$, i.e., $\bar\xi(x,a) = (a^2/a_i^2)\bar\xi(x,a_i)$.  In
addition we know that $\bar\xi \ll 1$ in this regime.  These can be
used, along with 
the pair conservation equation \cite{lssu} to show that
$h=2\bar\xi/3$ for $\bar\xi \ll 1$.  This equation is valid for all
models in linear theory.  We have plotted $h(a,x)$ as a function of
$\bar\xi(a,x)$ in 
figure~8.  We have plotted the line corresponding to the linear
theory result and have shown results from N-Body simulations as
points.  We used simulations of power law models [$n=0$ and $-1$] and
the standard CDM model for this plot.  Points corresponding to
different simulations have been shown with different symbols.  All the
points crowd around the line $h=2\bar\xi/3$ in the linear regime with
little dispersion.  The points deviate from this line for large
$\bar\xi$ but the dispersion between different models remains small.
This corresponds to an approximate ``universality'' in the relation
between $h$ and $\bar\xi$. [\cite{hamilref}; \cite{rntp}]

\begin{figure}
\begin{center}
\includegraphics[width=5truein]{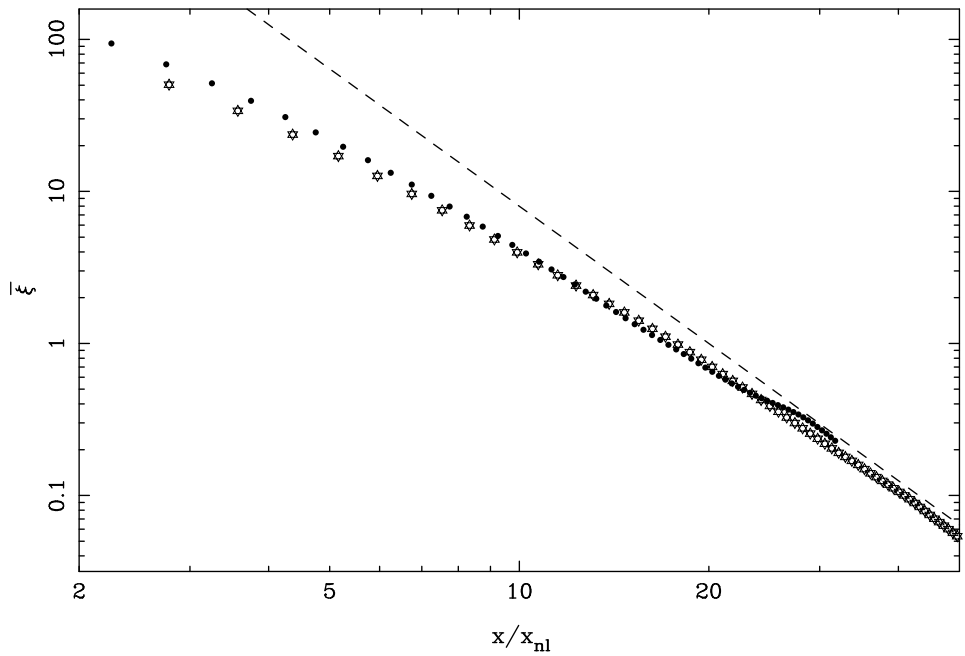}
\end{center}
\caption{This figure shows $\bar\xi(a,x)$ as a function of $x/x_{nl}$
for the $n=0$ power spectrum.  The points have been plotted for two
epochs.  The overlap between points clearly shows that the system is
evolving in a self similar manner.  The dashed line shows the linear
slope of the averaged correlation function.}
\end{figure}

Self similar evolution of $\bar\xi$ for power law models (\cite{lssu})
can be used to test correctness of the non-linear evolution of
gravitational clustering.  In figure~9 we have plotted
$\bar\xi(a,x)$ as a function of $x/x_{nl}$ for the $n=0$ power law
model.  Here $x_{nl}$ is the scale where the linearly extrapolated
$\bar\xi$ is unity.  This scale is proportional to $a^{2/3}$ for the $n=0$
power law model.  If the evolution is self-similar then the points from
different epochs must lie along a single curve.  Figure~9 clearly
shows that the evolution of averaged correlation function follows a
self similar pattern over a large range of scales. 

\begin{figure}
\begin{center}
\includegraphics[width=5truein]{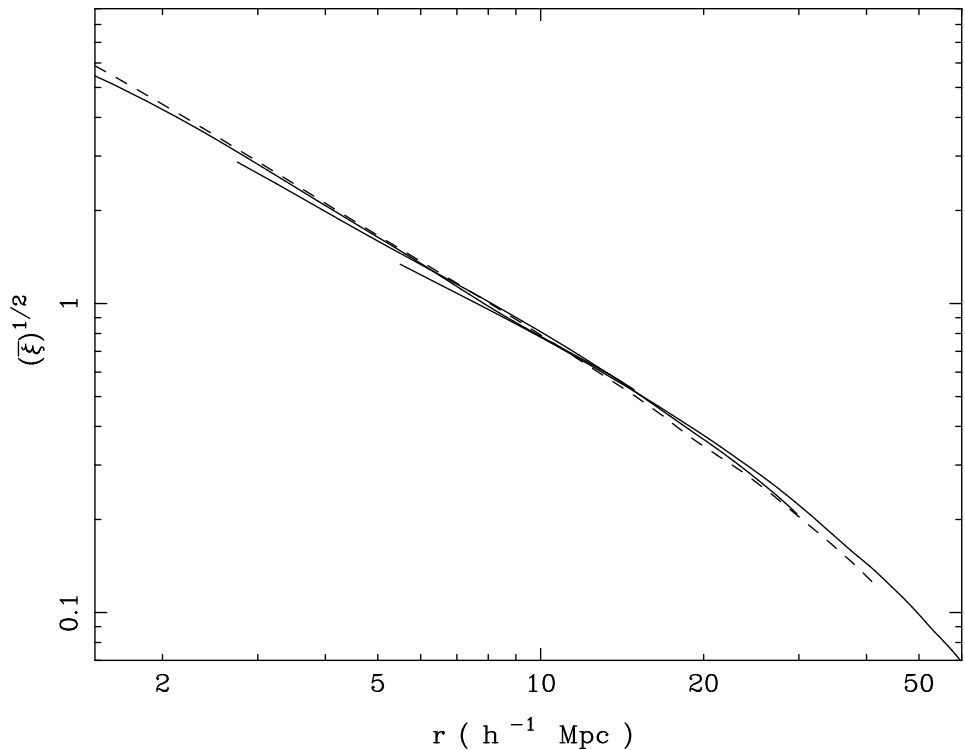}
\end{center}
\caption{Comparison of simulations at different scales.  This figure
shows $\bar\xi^{1/2}$ for three CDM simulations done with boxsize of $64$,
$128$ and $256h^{-1}$Mpc.  Matching of curves in the overlapping
region shows that the numerical simulation is not introducing any
artifacts.  The dashed line shows the same function from a P$^3$M
simulation by Brieur et al.(1995).  This simulation was done with the same
initial power spectrum but a different realisation of the Gaussian
random field was used.  The similarity between $\bar\xi$ in the 
three PM simulations and the P$^3$M simulation shows that the
non-linear evolution is being followed in the same manner in both the
simulations.} 
\end{figure}

The last test we consider here compares the averaged correlation
function for simulations of the standard CDM model.  We compare
$\bar\xi^{1/2}$ for three simulations carried out with boxsize $64$, $128$
and $256h^{-1}$Mpc.  Figure~10 shows that the curves match in the
overlapping region, implying that the numerical simulation is not
introducing any scale in the non-linear evolution of density
perturbations.  We also compare these curves with $\bar\xi^{1/2}$ obtained
from a different simulation.  This reference simulation was done by
Brieur, Summers and Ostriker (1995) using a P$^3$M code implemented on
a GRAPE machine.  This 
simulation was done with the same theoretical power spectrum but using
a different realisation of the initial density field.  The two types
of simulations being compared have virtually nothing in common as far
as the implementation of the mathematical model is concerned.  The
similarity in these curves implies that both the codes are evolving 
density perturbations in the same manner.

\section{Analysis of N-Body Data}

In this section we will outline the method used for computing the
correlation function and the scaled pair velocity (\cite{lssu}) from
simulation data.

The averaged pair velocity $h$ is defined as 
\begin{eqnarray}
h(a,x)&=& \left\langle \frac{a \left( \frac{d{\bf x}_i}{da} - \frac{d{\bf
x}_j}{da} \right).\left({\bf x}_i - {\bf x}_j \right)}{\left({\bf x}_i -
{\bf x}_j \right)^2} \right\rangle_{i \neq j} \nonumber\\
&=& \frac{\sum\limits_{i
\leq j}\frac{a \left( \frac{d{\bf x}_i}{da} - \frac{d{\bf x}_j}{da}
\right).\left({\bf x}_i - {\bf x}_j \right)}{\left({\bf x}_i - {\bf x}_i
\right)^2}}{\sum\limits_{i \leq j}  1}
\end{eqnarray}
In this equation subscripts label particles and the averaging is over
all pairs with separation equal to $x$, i.e., $x^2 = ({\bf x}_i - {\bf
x}_j )^2$.  This quantity is computed from simulation data by binning
the pairs by pair separation $x$.  The number of pairs in a given bin
as well as the quantity to be averaged is summed for each pair in each
bin.  The ratio of these quantities for each bin gives the value of
$h(a,x)$. 

The correlation function was defined in \S{1.4} as the Fourier transform
of the power spectrum.  This is equivalent to the following definition
\begin{equation}
\xi(a,x)=\langle \delta({\bf y})\delta({\bf z}) \rangle  \;\;\; ;
x=|{\bf y} - {\bf z}|
\end{equation}
where the average is over all pairs of points with separation ${\bf x}$
{\it and over all ${\bf x}$ with the same magnitude.}  This can be
rewritten as
\begin{eqnarray}
\xi(a,x)&=&\left\langle\left(\frac{\varrho(a,{\bf y})}{\bar\varrho(a)} - 1
\right)\left(\frac{\varrho(a,{\bf z})}{\bar\varrho(a)} - 1 \right)
\right\rangle \nonumber\\
&=& \left\langle\left(\frac{\varrho(a,{\bf
y})}{\bar\varrho(a)}\frac{\varrho(a,{\bf z})}{\bar\varrho(a)}\right) - 1
\right\rangle
\end{eqnarray}
Here we have used the fact that density contrast is a random field with
zero mean.  We can replace the
densities by number densities and the product of number densities
at points separated by a given distance by the number of pairs with that
separation.  With this the above equation reduces to
\begin{equation}
\xi(a,x)= \frac{n(a,x)}{\bar{n}(a)} - 1
\end{equation}
where $n(a,x)$ is the number of pairs with separation $x$ and $\bar n$ is
the number of pairs in a uniform distribution.  Therefore the problem is
again reduced to computing the number of pairs separated by a given
distance.  This is done by dividing the range of $x$ into small intervals
and binning the pairs into these intervals.  Similar operation is
required for computing the scaled pair velocity, therefore the binning
operation for computing these quantities can be combined conveniently.

We can obtain an expression for the averaged correlation by averaging
$\xi$.
\begin{eqnarray}
\bar\xi(a,x)&=&\frac{3}{x^3}\int\limits^x dy y^2 \xi(a,y) =
\frac{\int\limits^x dy y^2 \xi(a,y)}{\int\limits^x dy y^2} \nonumber\\
&=& \frac{\int\limits^x dy n(y)}{\int\limits^x dy \bar n(y)} - 1
\end{eqnarray}
where we have used the fact that the average number of pairs with a given
separation is proportional to the square of the separation for a uniform
distribution of particles.  In the discrete realisation of this
expression the integrals over number of pairs are replaced by summation
over bins.  Computationally this quantity is less error prone than the
correlation function as for large separations the number of pairs in a
given bin in the
simulation output approaches the number of pairs for a uniform
distribution.  Subtracting two nearly equal numbers can give large
error.  However, in case of the averaged correlation function the excess
number of pairs at small scales is carried over to large scales through
summation over all smaller scales.  And in general the ratio $(\sum
n)/(\sum\bar n)$ is larger at all scales in comparison with $n/\bar
n$.

\section{Discussion}

In the preceding sections we have described the basic mathematical
model that is implemented in particle mesh cosmological N-Body codes.
We have not discussed the P$^3$M codes, tree codes, etc. in this
review as only PM codes are known to ensure collisionless evolution
(\cite{plane}) though they suffer from a very limited resolution.  
Other methods improve spatial resolution but do not ensure
collisionless evolution.  However, most of the the machinery is common
to these codes and many of the results can be carried over to these
codes. 

\section*{ACKNOWLEDGEMENTS}

Authors thank S.F.Shandarin and F.R.Bouchet for useful
discussions on many aspects of N-Body codes.  JSB thanks CSIR India
for the Senior Research Fellowship.

\label{lastpage}

\end{document}